\documentclass[aps, pra, 
twocolumn, 
amsmath, amssymb, showpacs, 
]{revtex4-1}

\usepackage[utf8]{inputenc}
\usepackage{graphicx}
\usepackage{units}
\usepackage{color}
\usepackage[pdftex, colorlinks=true, linkcolor=myblue, citecolor=myblue,
urlcolor=myblue]{hyperref}
\usepackage{times}
\usepackage{bbold}
\usepackage{comment}

\definecolor{myblue}{rgb}{0,0,1}

\newcommand{\IPCMS}{Institut de Physique et Chimie des Mat\'{e}riaux de Strasbourg,
Universit\'{e} de Strasbourg, CNRS UMR 7504, F-67034 Strasbourg, France}

\begin{document}

\title{Nonradiative limitations to plasmon propagation in chains of metallic nanoparticles}

\author{Adam Brandstetter-Kunc}
\affiliation{\IPCMS}

\author{Guillaume Weick}
\email{guillaume.weick@ipcms.unistra.fr}
\affiliation{\IPCMS}

\author{Charles A. Downing}
\affiliation{\IPCMS}

\author{Dietmar Weinmann}
\affiliation{\IPCMS}

\author{Rodolfo A.\ Jalabert}
\affiliation{\IPCMS}


\begin{abstract}
We investigate the collective plasmonic modes in a chain of metallic nanoparticles that are 
coupled by near-field interactions. 
The size- and momentum-dependent 
nonradiative Landau damping and radiative decay rates are calculated 
analytically within an open quantum system approach. 
These decay rates determine the excitation propagation along the 
chain. In particular, the behavior of the radiative decay rate as a function 
of the plasmon wavelength leads to a transition from an exponential decay 
of the collective excitation for short distances to an algebraic decay for large distances. 
Importantly, we show that the exponential decay is of a purely nonradiative origin. 
Our transparent model enables us to provide analytical expressions for the polarization-dependent 
plasmon excitation profile along the chain and for the associated propagation length.  
Our theoretical analysis constitutes an important 
step in the quest for the optimal conditions for plasmonic propagation in nanoparticle chains.
\end{abstract}

\pacs{73.20.Mf, 73.22.Lp, 78.67.Bf}

\maketitle

\section{Introduction}

One of the primary goals of plasmonics \cite{maier} is to confine light at 
subwavelength scales in order to transport and manipulate it over macroscopic 
distances. While metallic nanostructures have been proposed and widely studied 
to achieve such ``plasmonic circuits" \cite{barne03_Nature}, both radiative 
and nonradiative losses inherent to metals are rather significant and, hence, 
limit the possible applications for energy and information transport at the nanoscale \cite{khurg15_NatureNanotech}. 
Thus, 
understanding the different damping mechanisms in 
radiatively coupled metallic nanostructures is of paramount 
interest in the field of plasmonics from a fundamental point of view and in 
order to increase the efficiency of signal transmission. 

The proposal of using a linear chain of spherical metallic nanoparticles as a 
subwavelength-sized light guide \cite{Quin98_OL} was accompanied by classical 
electromagnetic calculations based on the generalized Mie 
theory \cite{gerar82_PRB}.
Solving Maxwell's equations for a driven chain of Ag nanoparticles with 
radius $a=\unit[25]{nm}$ (the incoming light illuminating only the first 
nanoparticle of the chain), 
the authors of Ref.\ \cite{Quin98_OL} attempted to optimize the interparticle 
distance to achieve maximum propagation length. Significant propagation 
was only found for the longitudinal excitation
(with the electric field parallel to the axis of the chain). The largest 
propagation length ($\unit[900]{nm}$) was obtained for a center-to-center 
interparticle distance 
of $d=3a$. 

Plasmonic chains, as well as structures containing corners and 
junctions, were studied by Brongersma \textit{et al}.\ \cite{Brog00_PRB} within 
a model description based on electrostatically coupled point dipoles. An 
analytic form of the dispersion relation for the longitudinal and transverse
modes  was given and shown to be weakly affected by couplings beyond nearest 
neighbors. 
Assuming the radiation damping to be the same as that of uncoupled nanoparticles, 
the authors of Ref.\ \cite{Brog00_PRB} found a negligible radiation damping  
based on an estimation of the radiation from a single oscillating electron.
Under these two questionable assumptions, 
the dominant losses were of nonradiative character (phonons, electrons, lattice defects and impurities), leading to 
comparable attenuations of the longitudinal and 
transverse modes, and similar propagation lengths to those found in 
Ref.~\cite{Quin98_OL}.
The dispersion relation obtained using 
the point dipole model was later validated by using finite-difference time-domain calculations for an Au nanoparticle chain 
\cite{meier_PRB03}.
The possibility of propagating a pulse excitation was confirmed in 
Ref.\ \cite{meier_PRB03}, while the attenuation 
($\unit[3]{dB}/\unit[140]{nm}$ and $\unit[3]{dB}/\unit[43]{nm}$ for the 
longitudinal and transverse modes, respectively) was larger than that 
obtained in Ref.\ \cite{Brog00_PRB} for Ag nanoparticle chains.

Further theoretical studies investigated the influence of retardation effects 
in the dipole-dipole interaction on the plasmonic properties of the chain. It 
was found \cite{weber_PRB04, citrin_NL04} that a nonmonotonic 
behavior in the dispersion relation of the transverse mode emerges due to 
retardation effects, unlike the case of the longitudinal mode.
This non-monotonic behavior was argued to arise from
the phase matching of the 
plasmon dispersion with free photons of the same frequency \cite{weber_PRB04}.
Moreover, the influence of the interaction 
on radiation losses was shown to give a mode-dependent radiation damping 
\cite{weber_PRB04, citrin_NL04}.
The decay of the plasmon propagation in a driven chain was found to be 
non-exponential for both the longitudinal and transverse modes 
\cite{weber_PRB04}, with the transverse mode persisting for longer 
distances than the longitudinal one.

Later studies using a similar retarded approach considered ordered and 
disordered chains of metallic nanoparticles \cite{markel_PRB07}. 
While a similar behavior for the dispersion and radiation damping to that 
found in Ref.\ \cite{weber_PRB04} was observed, a distinction between two 
types of plasmons was introduced: ordinary, subradiative modes that localize 
in the presence of any disorder strength, and extraordinary, radiative modes 
that depend weakly on disorder.
In the past few years, the quantum properties of metallic nanoparticle chains has
also attracted some attention, as such chains may serve as quantum 
communication devices \cite{Lee_PRB12} and they might present significant 
entanglement stored in the collective modes \cite{Pino_PRL14}.

On the experimental side, the first observation of the near field associated with 
collective plasmons in ordered nanoparticle chains was reported by Krenn 
and coworkers \cite{krenn_PRL99} using a photon scanning tunneling 
microscope. 
The near-field optical effects measured with a scanning tunneling microscope for 
a chain of 10000 half-oblate spheroidal Au nanoparticles of
dimensions $100\times100\times\unit[40]{nm^3}$, separated by a 
distance of $\unit[100]{nm}$, were found to be consistent with the numerical 
solution of Maxwell's equations. 
Using a far-field spectroscopy technique, 
Maier \textit{et al}.\ \cite{maier_PRB02} measured the frequencies of the 
infinite-wavelength longitudinal and transverse modes for a chain composed of 
80 almost spherical Au nanoparticles with $a=\unit[25]{nm}$ and $d=75$, $100$ and $\unit[125]{nm}$. These frequencies were 
found to be in agreement with
the predictions of Ref.\ \cite{Brog00_PRB}, and in particular with their $d$-dependence.

Later experimental studies \cite{maier_Nat03} used a near-field scanning 
optical microscope and fluorescent dyes to investigate energy 
transport along a chain of Ag nanoparticles with sizes 
$90\times30\times\unit[30]{nm^3}$, separated 
by a distance of $\unit[50]{nm}$, and they obtained an attenuation of the plasmon 
excitation of $\unit[6]{dB}$ over $\unit[195]{nm}$.
The recent development of experimental techniques 
allowed for a spatial imaging of the electric field associated with the 
plasmons along a chain of elliptical nanocylinders. 
\cite{Apuzzo_NL13}. Furthermore, the electron energy loss spectroscopy 
technique was used to excite and map subradiant modes of short nanoparticle 
chains \cite{Barrow_NL14}.

While the existence of collective plasmons in nanoparticle chains and the theoretical approaches
predicting the resulting frequencies are well documented in the literature, the situation is more 
controversial when faced with the issue of the damping mechanisms, which are crucial for the 
excitation propagation and the practical application of these plasmonic waveguides.  
In this paper we investigate theoretically the problem of collective 
plasmonic excitations in chains of metallic nanoparticles, focusing on their damping mechanisms. 
We assume that the dipolar localized surface plasmons (LSPs) supported by 
each spherical nanoparticle interact through their near field and hence form 
plasmonic modes that are delocalized over the whole chain, i.e., 
collective plasmons. 
These collective excitations, like the LSP in single nanoparticles, suffer from both radiative and 
nonradiative losses. The former arise from the coupling between the collective modes and the 
photonic environment. The latter stem from Ohmic (absorption) losses characteristic of the bulk metal, 
and the coupling of the plasmon to electron-hole pairs, leading to a size-dependent Landau damping. 

We derive 
analytical expressions for the radiative damping rates of the transverse and 
longitudinal plasmonic modes in the infinite chain limit, confirming previous 
numerical studies. 
Our open quantum system approach further enables us to provide analytical expressions of the Landau 
damping decay rates, the latter being crucial for small nanoparticle sizes 
and/or for dark modes that couple only weakly to photons. 

Our approach based on the collective plasmon reduced density 
matrix in momentum space allows us to study energy transport along the chain. 
Importantly, we find that radiation damping is responsible for changing the character of the 
collective plasmon decay along the chain. 
While without radiation damping, the decay is exponential for all distances, the presence of radiation 
damping induces algebraic tails at long distances.
Such behavior is crucial for the appropriate characterization of the 
damping in the propagation of an initially localized excitation. 
Notably, we demonstrate that the short-distance exponential decay, which is the most relevant in 
the prospect of light and energy transport at the nanoscale, is of purely \textit{nonradiative} origin. 
We further show that the size-dependent Landau damping is crucial in understanding the limiting 
mechanisms to plasmon propagation, especially for small nanoparticles. 
Moreover, we provide analytical expressions for the plasmon excitation profile along the chain as well 
as for the associated polarization-dependent propagation length, which both reproduce numerical calculations 
with excellent agreement. 

The paper is organized as follows: Section \ref{sec:model} presents an open 
quantum system model to plasmon propagation in metallic nanoparticle chains. 
In Sec.\ \ref{sec:gamma}, we derive analytical expressions for both the 
Landau damping and radiative linewidths of the collective plasmons. The 
plasmon propagation along the chain is studied both numerically and analytically 
in Sec.\ \ref{sec:propagation} before we conclude in Sec.\ \ref{sec:ccl}. We provide in Appendix 
\ref{sec:RWA} a discussion of the rotating wave approximation for the plasmon 
dynamics, and in Appendix \ref{sec:dimer} we provide a detailed 
analysis of the special case of a heterogeneous nanoparticle dimer.
We relegate to Appendix \ref{sec:integrals} a few mathematical details for the derivation 
of the plasmon propagation length along the nanoparticle chain.

\section{Model}
\label{sec:model}

\begin{figure}[b]
 \includegraphics[width=\columnwidth]{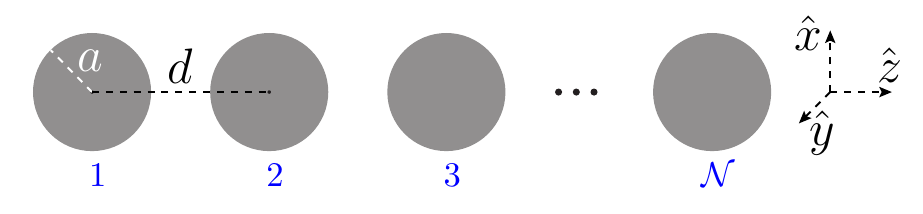}
\caption{Sketch of a linear chain of $\mathcal{N}$ identical spherical metallic nanoparticles.}
 \label{fig:sys}
 \end{figure}

We consider a linear chain of $\mathcal{N}$ identical spherical metallic 
nanoparticles of radius $a$ separated by a distance $d$ as sketched in 
Fig.~\ref{fig:sys}.
Each nanoparticle can sustain three degenerate LSP resonances that couple to the 
neighboring ones via the quasi-static dipole-dipole interaction 
\cite{maier,Brog00_PRB,park_PRB04}.
As in the cases of a single metallic nanoparticle \cite{gerch02_PRA,prb2006} 
and a nanoparticle dimer \cite{my}, separating the electronic coordinates into 
center-of-mass and relative coordinates 
yields a description typical for an open quantum system. The dipolar LSPs 
(the center-of-mass coordinates of the electron gas) are coupled
to electronic environments (baths of electron-hole pairs 
represented by the relative coordinates) present
in each nanoparticle. This nonradiative mechanism leads to the Landau damping 
\cite{bertsch,Kawa66_JPS,Yalu92_PRA,prb2005,prb2006,my} of the collective 
excitations, a purely quantum-mechanical effect. The coupling between 
the plasmonic and electron-hole pair subsystems is a consequence of the 
breaking of Kohn's theorem \cite{kohn61_PR,jacak} due to the nonharmonicity of 
the single-electron confinement arising from the positive ionic background 
\cite{gerch02_PRA,prb2006}.
In addition, the LSPs couple to the electromagnetic field modes, leading to the radiative decay of the collective plasmons.
As previously stated, Ohmic losses inherent to the bulk metal provide a further nonradiative decay channel for the collective modes.

\subsection{Hamiltonian of the system}
We write the Hamiltonian of the system as
\begin{equation}
\label{eq:H}
 H=H_\mathrm{pl}+H_\mathrm{eh}+H_\mathrm{ph}+H_\mathrm{pl\textrm{-}eh}+H_\mathrm{pl\textrm{-}ph} +H_\mathrm{drive},
\end{equation}
where the plasmonic part reads 
\begin{align}
 \label{eq:Hpl}
 H_\mathrm{pl}=&\;
 \hbar\omega_0\sum_{n=1}^\mathcal{N}\sum_{\sigma=x,y,z} 
 {b_{n}^{\sigma}}^\dagger b_{n}^{\sigma}
 \nonumber\\
 &+
 \hbar\Omega\sum_{n=1}^{\mathcal{N}-1}\sum_{\sigma=x,y,z} 
\eta_\sigma
\left(b_{n}^{\sigma}+ {b_{n}^{\sigma}}^\dagger\right)\left(b_{n+1}^{\sigma}+ {b_{n+1}^{\sigma}}^\dagger\right)
\end{align}
with $n$ the index identifying the particle number in the chain (see Fig.\ 
\ref{fig:sys}). Here, $\eta_x=\eta_y=1$ for the two transverse 
polarizations and $\eta_z=-2$ for the longitudinal one. Each nanoparticle 
supports three degenerate dipolar LSPs with a resonance frequency $\omega_0$ 
that, for alkaline nanoparticles in vacuum and neglecting the spill-out 
effect \cite{kreibig}, corresponds to the Mie frequency 
$\omega_\mathrm{p}/3^{1/2}=(N_\mathrm{e}e^2/m_\mathrm{e}a^3)^{1/2}$, 
where $\omega_\mathrm{p}=(4\pi n_\mathrm{e}e^2/m_\mathrm{e})^{1/2}$ is the plasma frequency 
of the considered metal.
Here, $e$ denotes the electron charge, $m_\mathrm{e}$ is its mass, 
$N_\mathrm{e}$ is the number of electrons in each nanoparticle, and $n_\mathrm{e}$ is the corresponding electronic density. 
The bosonic operator $b_{n}^{\sigma}$ ($ {b_{n}^{\sigma}}^\dagger$) in Eq.\ 
\eqref{eq:Hpl}
annihilates (creates) an LSP in the $\sigma=x,y,z$ direction in the 
$n$\textsuperscript{th} nanoparticle. 
In the regime $3a\lesssim d \ll k_0^{-1}$ \cite{park_PRB04}, where $k_0=\omega_0/c$ is the 
wavenumber corresponding to the LSP frequency ($c$ is the speed of light in vacuum), 
the LSPs couple to their nearest neighbors 
essentially via the near-field quasistatic dipole-dipole interaction. The latter
gives rise to the second term on the right-hand side of Eq.\ \eqref{eq:Hpl} 
\cite{Weick_PRL,Sturges_2DMat,epjb2015,my, lamow16_preprint}
with the coupling constant 
\begin{equation}
\label{eq:Omega}
\Omega=\frac{\omega_0}{2} \left(\frac{a}{d}\right)^3. 
\end{equation}
We do not use the rotating wave approximation in Eq.\ \eqref{eq:Hpl}, since the 
nonresonant terms $\propto (b_{n}^{\sigma}b_{n+1}^{\sigma}+\mathrm{h.c.})$ 
are important for the plasmonic eigenstates, and hence for quantities derived 
from them, such as, e.g., plasmon lifetimes (for more details, see Appendix 
\ref{sec:RWA}).

Our open chain of coupled metallic nanoparticles is conveniently described using the basis
\begin{equation}
\label{eq:sinetransform}
b_n^\sigma=\sqrt{\frac{2}{\mathcal{N}+1}}\sum_q\sin{(nqd)}\, b_q^\sigma
\end{equation}
with $q={\pi m}/{(\mathcal{N}+1)d}$ the plasmonic momentum, where the integer $m\in [1,\mathcal{N}]$. 
Using Eq.\ \eqref{eq:sinetransform}, the plasmonic Hamiltonian \eqref{eq:Hpl} is expressed in momentum space, yielding
\begin{align}
H_\mathrm{pl}=&\sum_{q\sigma}\left[\hbar\omega_0+2\eta_\sigma\hbar\Omega\cos{(qd)}\right] {b_{q}^{\sigma}}^\dagger b_q^\sigma
\nonumber\\
&+\hbar\Omega\sum_{q\sigma}\eta_\sigma\cos{(qd)}\left({b_{q}^{\sigma}}^\dagger {b_{q}^{\sigma}}^\dagger+b_q^\sigma b_q^\sigma\right).
\end{align}
After diagonalization by means of a bosonic Bogoliubov transformation, the above Hamiltonian reads
\begin{equation}
 \label{eq:Hplq}
 H_\mathrm{pl}=\sum_{q\sigma}\hbar\omega_q^{\sigma} {B_{q}^{\sigma}}^\dagger B_q^{\sigma},
\end{equation}
where the eigenfrequencies of the collective plasmons are given by 
\begin{equation}
\label{eq:plasmon_spectrum}
\omega_q^\sigma=\omega_0\sqrt{1+4\eta_\sigma(\Omega/\omega_0)\cos{(qd)}}.
\end{equation}
In Eq.\ \eqref{eq:Hplq}, the bosonic operators
\begin{equation}
 \label{eq:bqs}
 B_q^{\sigma} =\cosh{(\theta^\sigma_q)}\, b_q^{\sigma}+\sinh{(\theta^\sigma_q)}\, {b_{q}^{\sigma}}^\dagger
\end{equation}
and their adjoints ${B_{q}^{\sigma}}^\dagger$ 
annihilate and create, respectively, a collective plasmon excitation with polarization $\sigma$ and with momentum $q$  along the chain. The coefficients of the Bogoliubov transformation in Eq.\ \eqref{eq:bqs} read
\begin{subequations}
\label{eq:Bogoliubov}
\begin{equation}
\cosh{(\theta_q^{\sigma})}=\frac{1}{\sqrt{2}}
\sqrt{\frac{1+2\eta_\sigma(\Omega/\omega_0)\cos(qd)}{\sqrt{1+4\eta_\sigma(\Omega/\omega_0)\cos(qd)}}+1}
\end{equation}
and
\begin{align}
\sinh{(\theta_q^{\sigma})}=&\; \frac{\mathrm{sgn}\left\{\eta_\sigma\cos(qd)\right\}}{\sqrt{2}}
\nonumber\\
&\times
\sqrt{\frac{1+2\eta_\sigma(\Omega/\omega_0)\cos(qd)}{\sqrt{1+4\eta_\sigma(\Omega/\omega_0)\cos(qd)}}-1}.
\end{align}
\end{subequations}

\begin{figure}[tb]
 \includegraphics[width=\columnwidth]{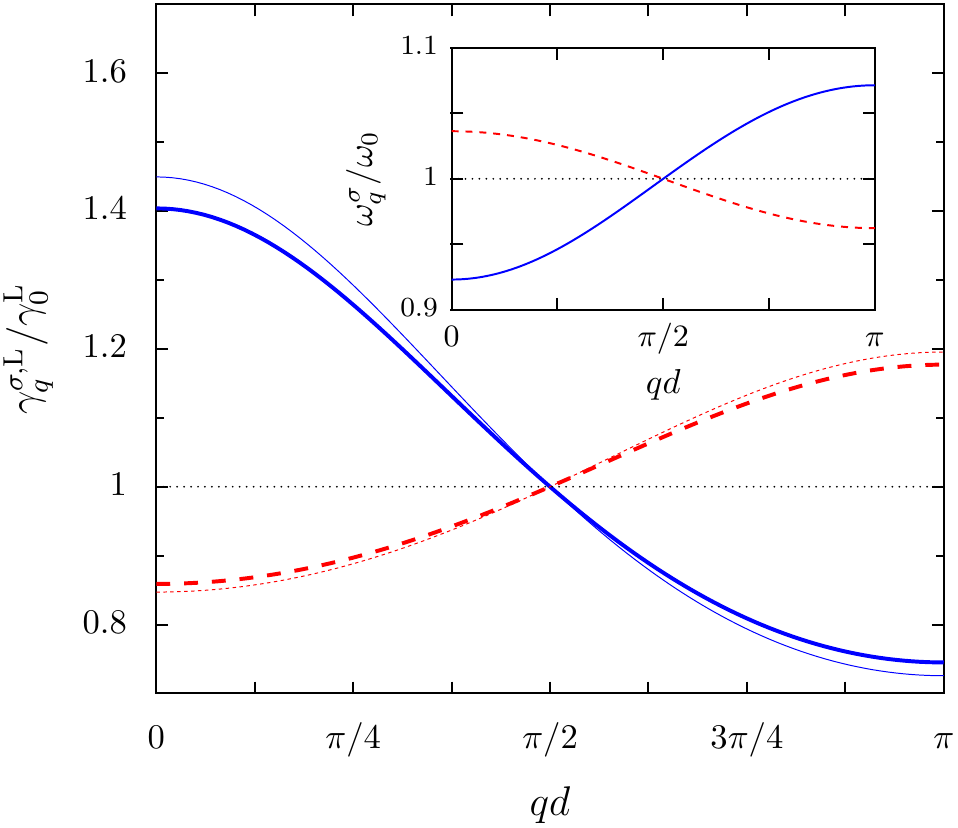}
\caption{Landau damping decay rate from Eq.\ \eqref{eq:landau} as a function of momentum for the transverse (red dashed lines, $\sigma=x,y$)
and longitudinal (blue solid lines, $\sigma=z$) collective plasmonic modes for an interparticle separation $d=3a$. The thick (thin) lines correspond to $\hbar\omega_0/E_\mathrm{F}=0.5$
 ($\hbar\omega_0/E_\mathrm{F}=1$). The inset shows the corresponding collective plasmon dispersions \eqref{eq:plasmon_spectrum}.
}
 \label{fig:rates}
 \end{figure}

We show in the inset of Fig.\ \ref{fig:rates} the dispersion relation \eqref{eq:plasmon_spectrum} of the transverse (red dashed line) and longitudinal (blue solid line) collective plasmons. 
Including the far-field corrections and the associated retardation effects in the dipole-dipole 
interaction between the nanoparticles along the chain, as was done in Ref.\ \cite{weber_PRB04}, only leads to
a slight quantitative modification with respect to the dispersion relations shown in the inset of Fig.\ \ref{fig:rates}. 
This justifies that we only consider the near-field interaction between nearest neighbors in
Eq.\ \eqref{eq:Hpl}. The dependence of the eigenfrequencies \eqref{eq:plasmon_spectrum} on the interparticle distance $d$ is encapsulated in the coupling constant 
$\Omega\ll\omega_0$ defined in Eq.\ \eqref{eq:Omega}, yielding $\omega_q^\sigma/\omega_0-1\simeq\eta_\sigma(a/d)^3\cos{(qd)}$. Such a $1/d^3$ dependence \cite{Brog00_PRB} directly stems from the scaling of the quasistatic dipole-dipole interaction with $d$.

The electronic environment is composed of electron-hole excitations and is described in Eq.\ \eqref{eq:H} by the Hamiltonian
\begin{equation}
 \label{eq:Heh}
 H_\mathrm{eh}=\sum_{n=1}^\mathcal{N} \sum_{\alpha}\varepsilon_{n\alpha}  c_{n\alpha}^\dagger c_{n\alpha}^{\phantom{\dagger}},
\end{equation}
where $c_{n\alpha}$ ($c_{n\alpha}^\dagger $) annihilates (creates) an electron 
in the $n$\textsuperscript{th} nanoparticle associated with the one-body 
state $\left|n\alpha\right>$
with energy $\varepsilon_{n\alpha}$ in the self-consistent potential $V_n$ of 
that nanoparticle. We assume $V_n$ to be a spherically 
symmetric hard-wall potential \cite{prb2006,zuloa09_NL,my}. 
The coupling of the plasmon to the electronic environment, arising from the 
nonharmonicity of the single-electron confinement, yields \cite{gerch02_PRA,prb2006,prb2005,my}
the coupling Hamiltonian $H_{\mathrm{pl\textrm{-}eh}}$ in Eq.\ \eqref{eq:H}
in the form
\begin{equation}
 \label{eq:Hpleh}
 H_\mathrm{pl\textrm{-}eh}=\Lambda \sum_{n=1}^\mathcal{N}  
 \sum_{\sigma=x,y,z}\sum_{\alpha\beta}\left(b_n^{\sigma}+{b_{n}^{\sigma}}^\dagger 
 \right)\langle n\alpha|\sigma|n\beta \rangle c_{n\alpha}^\dagger c_{n\beta}^{},
\end{equation}
with $\Lambda=(\hbar m_\mathrm{e}\omega_0^{3}/2N_\mathrm{e})^{1/2}$. 
The coupling Hamiltonian $H_\mathrm{pl\textrm{-}eh}$ is responsible for the Landau 
damping of the collective plasmons.

The plasmonic system is also coupled to a photonic bath described 
by the Hamiltonian
\begin{equation}
 \label{eq:Hph}
 H_\mathrm{ph}=
 \sum_{\mathbf{k},\hat{\lambda}_\mathbf{k}} \hbar\omega_{\mathbf{k}}
{a_{\mathbf{k}}^{\hat{\lambda}_\mathbf{k}}}^\dagger
a_{\mathbf{k}}^{\hat{\lambda}_\mathbf{k}},
\end{equation}
where $a_{\mathbf{k}}^{\hat{\lambda}_\mathbf{k}}$ 
(${a_{\mathbf{k}}^{\hat{\lambda}_\mathbf{k}}}^\dagger$) annihilates (creates)
a photon with momentum $\mathbf{k}$, transverse polarization $\hat{\lambda}_\mathbf{k}$ (i.e., $\hat{\lambda}_\mathbf{k}\cdot\mathbf{k}=0$),
and dispersion $\omega_\mathbf{k}= c|\mathbf{k}|$.
In the long-wavelength limit, 
assuming that the nanoparticle sizes are much smaller than $k_0^{-1}$,
the plasmon-photon coupling in Eq.~\eqref{eq:H} takes the form \cite{cohen}
\begin{equation}
\label{eq:minimal_coupling}
H_\mathrm{pl\textrm{-}ph}=\frac{e}{m_\mathrm{e}}\sum_{n=1}^\mathcal{N}
\boldsymbol{\Pi}_n\cdot\mathbf{A}(\mathbf{d}_n).
\end{equation} 
Here, 
\begin{equation}
\label{eq:LSP_momentum}
\boldsymbol{\Pi}_n=\mathrm{i}\sqrt{\frac{N_\mathrm{e}m_\mathrm{e}\hbar\omega_0}{2}}\sum_{\sigma=x,y,z}
\hat\sigma\left({b_n^\sigma}^\dagger-b_n^\sigma\right)
\end{equation}
is the momentum associated with the LSPs on nanoparticle $n$, 
and
\begin{equation}
\label{eq:vector_potential}
\mathbf{A}(\mathbf{d}_n)=\sum_{\mathbf{k},\hat{\lambda}_\mathbf{k}}
\hat{\lambda}_{\mathbf{k}}
\sqrt{\frac{2\pi\hbar}{\mathcal{V}\omega_\mathbf{k}}}
 \left(a_{\mathbf{k}}^{\hat{\lambda}_\mathbf{k}} \,
 \mathrm{e}^{\mathrm{i}\mathbf{k}\cdot \mathbf{d}_n}
 +{a_{\mathbf{k}}^{\hat{\lambda}_\mathbf{k}}}^\dagger\,
 \mathrm{e}^{-\mathrm{i}\mathbf{k}\cdot \mathbf{d}_n} \right)
\end{equation}
 is the vector potential evaluated at the position of the center of the $n^\textrm{th}$ nanoparticle $\mathbf{d}_n=\hat{z}(n-1)d$, where ${\mathcal{V}}$ is the quantization volume used for the electromagnetic modes. 
Together with Eqs.\ \eqref{eq:LSP_momentum} and \eqref{eq:vector_potential}, the plasmon-photon coupling 
\eqref{eq:minimal_coupling} thus takes the form
\begin{align}
 \label{eq:Hplph}
 H_\mathrm{pl\textrm{-}ph}=&\;
 \mathrm{i}\hbar\sum_{n=1}^\mathcal{N}\sum_{\sigma=x,y,z}\sum_{\mathbf{k},
 \hat{\lambda}_\mathbf{k}} 
 \sqrt{\frac{\pi\omega_0^3 a^3  }{{\mathcal{V}\omega_\mathbf{k}}}}
 \hat{\sigma}\cdot \hat{\lambda}_{\mathbf{k}}
 \nonumber\\
 &\times
 \left({b_{n}^{\sigma}}^\dagger-b_{n}^{\sigma}\right)
 \left(a_{\mathbf{k}}^{\hat{\lambda}_\mathbf{k}} \,
 \mathrm{e}^{\mathrm{i}\mathbf{k}\cdot \mathbf{d}_n}
 +{a_{\mathbf{k}}^{\hat{\lambda}_\mathbf{k}}}^\dagger\,
 \mathrm{e}^{-\mathrm{i}\mathbf{k}\cdot \mathbf{d}_n} \right).
\end{align}
 
The last term of the system Hamiltonian \eqref{eq:H} is a driving 
term representing an electric field, with wavelength much larger 
than the nanoparticle size, acting on the LSPs in the first nanoparticle. 
It reads
\begin{equation}
 \label{eq:drive}
H_\mathrm{drive}=\hbar\Omega_\mathrm{R}f(t)
\sum_{\sigma=x,y,z}\left( b_1^\sigma+{b_1^{\sigma}}^\dagger
\right)\hat{\sigma}\cdot\hat{\epsilon}
\end{equation}
with the Rabi frequency 
\begin{equation}
\label{eq:Rabi}
\Omega_\mathrm{R}=eE_0\sqrt{\frac{N_\mathrm{e}}{2m_\mathrm{e}\hbar\omega_0}},
\end{equation}
 where $E_0$ is the 
amplitude of the electric field and  
$\hat{\epsilon}$ its polarization. In Sec.\ \ref{sec:propagation}, we will consider both the case of 
a monochromatic electric field $f(t)=\sin{(\omega_\mathrm{d}t)}$ with $\omega_\mathrm{d}$ the driving frequency 
and the case of an extremely short laser pulse, modelled by $f(t)=\delta(\omega_0t)$, where $\delta(\nu)$ represents the Dirac delta function.

\subsection{Reduced density matrix}
The dynamics of the system is most conveniently described in terms of 
the reduced density matrix $\rho$ of the collective plasmonic degrees
of freedom. 
We treat the two coupling Hamiltonians $H_\mathrm{pl\textrm{-}eh}$ and 
$H_\mathrm{pl\textrm{-}ph}$ perturbatively and trace out the electronic and 
photonic degrees of freedom. 
The resulting time evolution at zero temperature in the large bath(s) limit and under the 
Markovian hypothesis is given by the Lindblad 
form~\cite{cohen, seoan07_EPJD,epjd2007, brand_unpublished}
\begin{align}
\label{eq:rho}
 \dot{\rho}=&
 -\mathrm{i}\sum_{q\sigma}
 \tilde{\omega}_q^\sigma\left[{B_q^{\sigma}}^\dagger B_q^{\sigma},\rho \right] 
 \nonumber\\
 &-\sum_{q\sigma}\frac{\gamma_q^\sigma}{2}
 \left( {B_q^{\sigma}}^\dagger B_q^{\sigma}\rho+\rho {B_q^{\sigma}}^\dagger B_q^{\sigma}
 -2B_q^{\sigma}\rho {B_q^{\sigma}}^\dagger\right)
\nonumber\\
&+\mathrm{i}\sum_{q\sigma}\frac{\mathcal{A}_q^\sigma f(t)}{2\tilde{\omega}_q^\sigma}
\left[{B_q^{\sigma}}^\dagger+ B_q^{\sigma},\rho \right],
\end{align}
where we introduced the amplitude of the driving term
\begin{equation}
\label{eq:A}
\mathcal{A}_q^\sigma=-2\sqrt{\frac{2}
{\mathcal{N}+1}}\hat{\sigma}\cdot\hat{\epsilon}\sin{(qd)}\, \Omega_\mathrm{R}
\tilde\omega_q^\sigma\sqrt{\frac{\omega_0}{\omega_q^\sigma}}.
\end{equation}

The rate $\gamma_q^\sigma=\gamma^\mathrm{O}+\gamma_q^{\sigma,
\mathrm{L}}+\gamma_q^{\sigma,\mathrm{r}}$ entering the master equation 
\eqref{eq:rho} and describing the decay
of a collective plasmonic mode $\{q,\sigma\}$ into the ground state consists of 
three components: (i) the nonradiative bulk Ohmic losses characterized by the (mode-independent) 
decay rate $\gamma^\mathrm{O}$ [which is phenomenologically incorporated in the master 
equation \eqref{eq:rho} and not through a first-principle calculation], (ii) the 
nonradiative Landau damping linewidth $\gamma_q^{\sigma,\mathrm{L}}$, and (iii) the radiative 
losses with decay rate $\gamma_q^{\sigma,\mathrm{r}}$. 

The Landau damping decay rate arising from the coupling 
Hamiltonian \eqref{eq:Hpleh} reads
\begin{equation}
 \label{eq:gammaL}
 \gamma_q^{\sigma,\mathrm{L}}=\frac{\omega_0}
 {\omega_q^\sigma}\Sigma^\sigma(\omega^\sigma_q)
\end{equation}
with
\begin{equation}
\label{eq:sigma}
\Sigma^\sigma(\omega)=\frac{2\pi}{\hbar^2}\Lambda^2\sum_{eh}\left|
\left<e\right|{\sigma} \left|h\right>\right|^2\delta(\omega-\omega_{eh}),
\end{equation}
where 
$\omega_{eh}=(\varepsilon_e-\varepsilon_h)/\hbar$, with $\left|e\right>$ and 
$\left|h\right>$ representing, respectively, electron and hole states with 
energy $\varepsilon_e$ and $\varepsilon_h$ in the self-consistent potential 
$V_n$
(assumed to be the same for each nanoparticle).
 
The radiative decay rate arising from the plasmon-photon interaction 
\eqref{eq:Hplph} is given by 
\begin{equation}
 \label{eq:gammaR}
 \gamma_q^{\sigma,\mathrm{r}}=2\pi^2\omega_0^2\omega_q^\sigma\frac{a^3}
 {\mathcal{V}}
 \sum_{\mathbf{k},\hat{\lambda}_\mathbf{k}}
\frac{{|\hat{\sigma}\cdot \hat{\lambda}_\mathbf{k}|}^2}{\omega_\mathbf{k}}
{|F_{\mathbf{k},q}|}^2 \delta(\omega^{\sigma}_q-\omega_\mathbf{k}),
\end{equation}
where the array factor
\begin{equation}
 \label{eq:F}
F_{\mathbf{k},q}=\sqrt{\frac{2}{\mathcal{N}+1}}\sum_{n=1}^\mathcal{N} 
\sin(nqd)\;\mathrm{e}^{-\mathrm{i}\mathbf{k}\cdot\mathbf{d}_n}
\end{equation}
is straightforwardly evaluated to yield
\begin{align} 
 \label{eq:AF}
 F_{\mathbf{k},q}=&\;\frac{\mathrm{i}\, \mathrm{e}^{\mathrm{i} k_z d}}
 {\sqrt{{2(\mathcal{N}+1)}}}
 \sum_{\kappa=\pm}\kappa\
 \mathrm{e}^{-\mathrm{i}\kappa(\mathcal{N}+1)(q+\kappa k_z)d/2} 
 \nonumber\\
 &\times
 \frac{\sin\left(
\mathcal{N} [q+\kappa k_z]d/2\right)}{\sin\left([q+\kappa k_q] 
d/2\right)}
 \end{align}
with $k_z$ referring to the $z$ component of the photon momentum $\mathbf{k}$.

In the master equation \eqref{eq:rho}, the eigenfrequency 
$\tilde{\omega}_q^\sigma=\omega_q^\sigma-(\delta_q^{\sigma,\rm 
L}+\delta_q^{\sigma,\rm r})$ contains the redshifts due 
to the interaction with electronic \cite{gerch02_PRA,prb2006, Hagino} and 
photonic \cite{cohen} environments, which read, respectively,
\begin{equation} 
 \label{eq:delqL}
\delta_q^{\sigma,\rm L}=\frac{2}{\hbar^2}\Lambda^2\frac{\omega_0}
{\omega_q^\sigma}\mathcal{P}\sum_{eh}|\langle e|{\sigma} |h\rangle|^2
\frac{\omega_{eh}}{\omega_{eh}^2-{\omega_q^\sigma}^2}
\end{equation}
 and 
 \begin{equation}
 \label{eq:delqr}
 \delta_q^{\sigma,\rm r}= 2\pi \omega_0^2\omega_q^\sigma\frac{a^3}
 {\mathcal{V}}\mathcal{P}
 \sum_{\mathbf{k},\hat{\lambda}_\mathbf{k}}
 |\hat{\sigma}\cdot\hat{\lambda}_\mathbf{k}|^2
 |F_{\mathbf{k}q}|^2
 \frac{1}{\omega_\mathbf{k}^2-{\omega_q^\sigma}^2}, 
 \end{equation}
where $\mathcal{P}$ denotes the Cauchy principal value.

\section{Nonradiative and radiative decay rates of the collective plasmonic 
modes}
\label{sec:gamma}
We now turn to the evaluation of the nonradiative and radiative decay rates 
given by the Fermi golden rule expressions in Eqs.\ \eqref{eq:gammaL} and 
\eqref{eq:gammaR}, respectively.

\subsection{Landau damping}
The function $\Sigma^\sigma(\omega)$, defined in Eq.\ \eqref{eq:sigma}, which determines the 
Landau damping \eqref{eq:gammaL} has been evaluated for $\sigma=z$ using 
semiclassical expansions \cite{prb2006}.
For symmetry reasons, $\Sigma^x(\omega)$ and $\Sigma^y(\omega)$ have the same 
expression as $\Sigma^z(\omega)$,
yielding in the zero-temperature limit [cf.\ Eq.\ (34) in Ref. \cite{prb2006}]
\begin{equation}
 \label{eq:landau}
 \gamma_q^{\sigma,\mathrm{L}}=\frac{3v_\mathrm{F}}{4a}\left(\frac{\omega_0}{\omega_q^\sigma}\right)^4g\left( \frac{\hbar\omega_q^\sigma}{E_\mathrm{F}} \right),
\end{equation}
where $v_\mathrm{F}$ and $E_\mathrm{F}$ are respectively the Fermi velocity and energy of the considered metal.
The function $g(\nu)$ entering the expression above is given by \cite{Yalu92_PRA,prb2011}
\begin{subequations}
\label{eq:g}
\begin{align}
g(\nu)=&\;
\frac{1}{3\nu}\left[(1+\nu)^{3/2}-(1-\nu)^{3/2}\right]
\nonumber\\
&+\frac{\nu}{4}\left(\sqrt{1+\nu}-\sqrt{1-\nu}-\nu\ln{\nu}\right)
\nonumber\\
&+\frac{\nu}{2}\left[
\left(1+\frac{\nu}{2}\right)\ln{\left(\sqrt{1+\nu}-1\right)}
\right.\nonumber\\
&-\left.\left(1-\frac{\nu}{2}\right)\ln{\left(1-\sqrt{1-\nu}\right)}
\right]
\end{align}
for $\nu\leqslant1$ and
\begin{align}
g(\nu)=&\;
\frac{1}{3\nu}(1+\nu)^{3/2}
+\frac{\nu}{4}\left(\sqrt{1+\nu}-\ln{\nu}\right)
\nonumber\\
&+\frac{\nu}{2}\left[
\left(1+\frac{\nu}{2}\right)\ln{\left(\sqrt{1+\nu}-1\right)}
-\frac{\nu}{2}\ln{\sqrt{\nu}}
\right]
\end{align}
\end{subequations}
for $\nu>1$.
The nonradiative decay rate \eqref{eq:landau} scales as the inverse of the 
nanoparticle size, so that for small enough nanoparticles,
Landau damping dominates over radiation damping (which scales as the particle 
volume; see Sec.\ \ref{sec:rad}). 
Landau damping is therefore of prominent importance in the prospect of light and 
energy transport in nanoscale plasmonic arrays. 

The Landau damping decay rates of the transverse and longitudinal collective 
plasmon modes are shown in Fig.\ \ref{fig:rates} as a function of momentum.
Once scaled with the Landau damping decay rate of a single nanoparticle 
\cite{Kawa66_JPS,Yalu92_PRA,prb2005} $\gamma_0^\mathrm{L}=(3v_\mathrm{F}/4a) 
g(\hbar\omega_0/E_\mathrm{F})$,
the nonradiative linewidths of the coupled plasmons show a significant 
modulation as a function of the wavelength of the modes (between $-25\%$ and 
$+45\%$ for the longitudinal mode and for $d=3a$). For larger interparticle distances, 
the modulation is less stringent, since for coupling constant $\Omega\ll\omega_0$, 
$\gamma_q^{\sigma, \mathrm{L}}/\gamma_0^\mathrm{L}-1\simeq
\eta_\sigma(a/d)^3\cos{(qd)}G(\hbar\omega_0/E_\mathrm{F})$ with $G(\nu)=\nu g'(\nu)/g(\nu)-4$, 
and where 
$g'(\nu)$ represents the derivative of the function $g(\nu)$ defined in Eq.\ \eqref{eq:g} with respect to $\nu$.
As can be seen in Fig.~\ref{fig:rates}, the higher the frequency of the mode, the 
lower is its Landau damping linewidth, similarly to the case of an isolated 
nanoparticle \cite{seoan07_EPJD}.
Notice also that the dependence of $\gamma_q^{\sigma,\mathrm{L}}$ on the ratio 
$\hbar\omega_0/E_\mathrm{F}$ is rather weak (thick and thin lines in 
Fig.~\ref{fig:rates} correspond, respectively, to $\hbar\omega_0/E_\mathrm{F}=0.5$ and $1$, values that are of the 
order of magnitude that is usually encountered in metallic nanoparticles). 
This is due to the relatively smooth behavior exhibited by the 
monotonically decreasing function $g(\nu)$.

\subsection{Radiative damping}
\label{sec:rad}

In the Fermi golden rule \eqref{eq:gammaR} for the radiative decay rate of the collective plasmons, the summation over photon polarizations is done using
 $\sum_{\hat{\lambda}_\mathbf{k}}|\hat{\sigma}\cdot\hat{\lambda}_\mathbf{k}|^2=1-(\hat{\sigma}\cdot\hat{k})^2$, while the sum over photonic momenta $\mathbf{k}$
is performed in the continuous limit ($\mathcal{V}\rightarrow\infty$). Using spherical coordinates for the integral over $\mathbf{k}$, we arrive at
\begin{align}
 \label{eq:grsigma}
\gamma_q^{\sigma,\mathrm{r}}=&\;
\frac{3|\eta_\sigma|\gamma_0^\mathrm{r}}{8}\left(\frac{\omega_q^{\sigma}}{\omega_0} \right)^2
\nonumber\\
&\times\int_0^\pi \mathrm{d}\theta \sin{\theta}\left(1+\mathrm{sgn}\{\eta_\sigma\}\cos^2\theta\right)|F_{\mathbf{k}_q^\sigma,q}|^2
\end{align}
for the radiative decay rates of the transverse and longitudinal 
collective plasmons. In Eq.~\eqref{eq:grsigma}, 
$\gamma_0^\mathrm{r}=2\omega_0^4a^3/3c^3$
is the radiation damping decay rate of a single isolated nanoparticle, and
\begin{align}
 \label{eq:fq2}
|F_{\mathbf{k}_q^\sigma,q}|^2=&\;\frac{1}{2(\mathcal{N}+1)}
\left\{\sum_{\kappa=\pm}\frac{\sin^2\left(\mathcal{N}[q+\kappa k_q^\sigma 
\cos\theta]d/2\right)}{\sin^2\left([q+\kappa k_q^\sigma\cos\theta]d/2 \right)}\right.
\nonumber\\
& - 2\cos([\mathcal{N}+1]qd)
\nonumber\\
&\left.\times
\prod_{\kappa=\pm}
\frac{\sin\left(\mathcal{N}[q+\kappa k_q^\sigma\cos\theta]d/2\right)}
{\sin\left([q+\kappa k_q^\sigma\cos\theta]d/2\right)}  \right\},
\end{align}
where $k_q^\sigma=\omega_q^\sigma/c$. In the infinite chain limit 
($\mathcal{N}\rightarrow\infty$), the expression above reduces to 
\begin{equation}
|F_{\mathbf{k}_q^\sigma,q}|^2\simeq\pi\sum_{\kappa=\pm}
\delta([q+\kappa k_q^\sigma\cos\theta]d),
\end{equation}
 such that the remaining integral in Eq.\ \eqref{eq:grsigma} is easily performed, and yields
\begin{equation}
 \label{eq:radiationsigma}
 \gamma_q^{\sigma,\mathrm{r}}=
 \frac{3\pi|\eta_\sigma|\gamma_0^\mathrm{r}}{4k_0 d}\,
 \frac{\big(\omega_q^{\sigma}\big)^2+\mathrm{sgn}\{\eta_\sigma\}\big(cq\big)^2}
 {\omega_0\omega_q^{\sigma}}\, \Theta\left(\omega_q^{\sigma}-cq\right).
\end{equation}
We denote by $\Theta(\nu)$ the Heaviside step function.

\begin{figure}[tb]
 \includegraphics[width=\columnwidth]{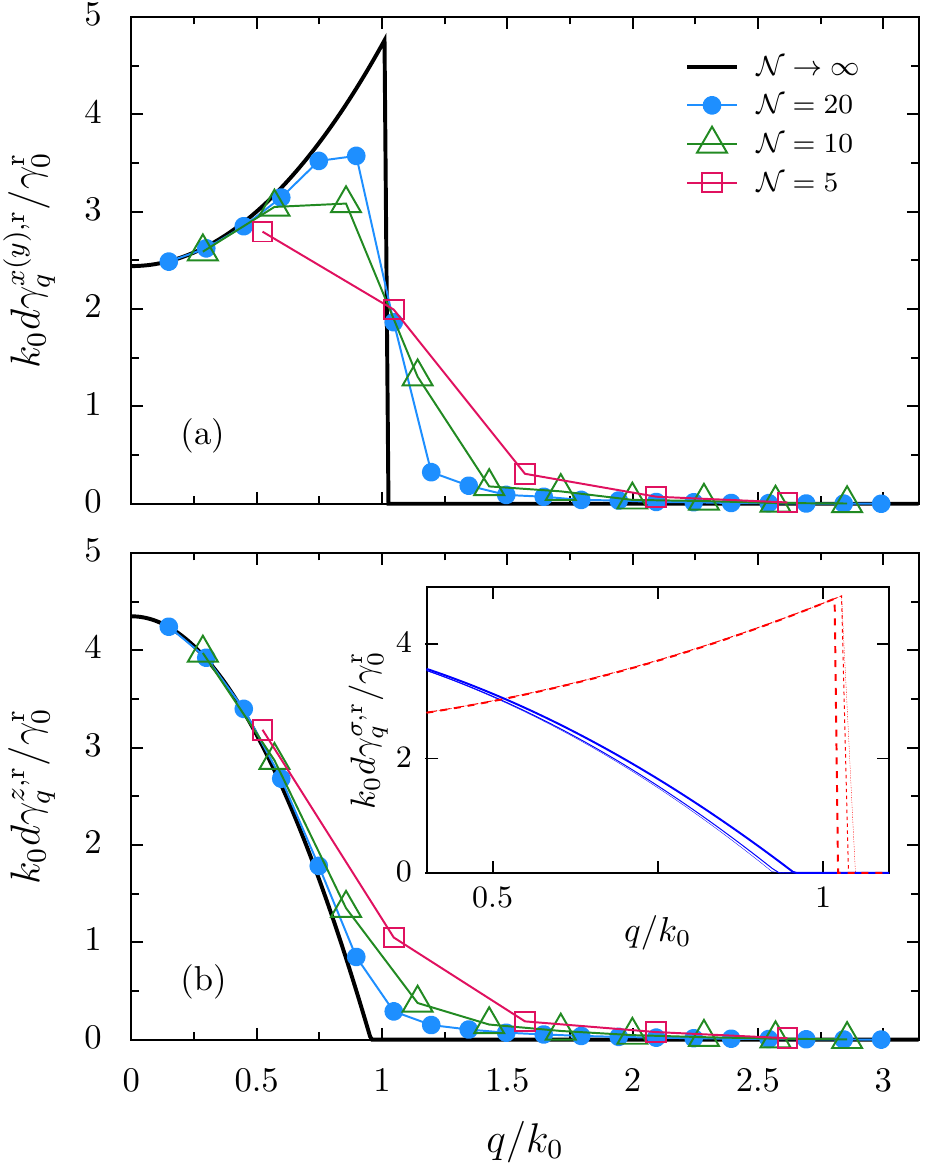}
\caption{Radiation damping decay rate from Eq.\ \eqref{eq:grsigma} as a 
function of momentum for the (a) transverse and (b) longitudinal collective plasmonic 
modes for $k_0d=1$ in chains with $d=3a$ that contain various numbers $\mathcal{N}$ of 
nanoparticles. The thick solid lines correspond to 
$\mathcal{N}\rightarrow\infty$, [cf.\  Eq.\ \eqref{eq:radiationsigma}]. 
The inset shows the radiation damping decay rate from Eq.\ \eqref{eq:radiationsigma} 
for the transverse (red dashed lines) and 
longitudinal (blue solid lines) collective plasmonic modes for 
$k_0d=0.25$, $0.5$, and $1$ from the thin to the thicker line.
}\label{fig:rad}
\end{figure}

In Fig. \ref{fig:rad}, we compare our analytical results for transverse [Fig.\ \ref{fig:rad}(a)] 
and longitudinal [Fig.\ \ref{fig:rad}(b)] plasmonic modes in infinite chains 
($\mathcal{N}\rightarrow \infty$), Eq.\ \eqref{eq:radiationsigma}, to a numerical
evaluation of Eq.~\eqref{eq:grsigma} for finite chains containing 
$\mathcal{N}=\{5,10,20\}$ nanoparticles.
The special case of a nanoparticle dimer ($\mathcal{N}=2$), where the momentum representation is of no use and which has 
already been considered in Ref.\ \cite{my}, is presented in Appendix \ref{sec:dimer}.
As one can see from Fig.\ \ref{fig:rad}, the behavior of the finite chain approaches 
the analytical infinite chain limit with rather good agreement already
for $\mathcal{N}=20$. 
For $\mathcal{N}=50$, the continuous black line representing the analytical 
result of Eq.\ \eqref{eq:radiationsigma} in Fig.\ \ref{fig:rad} and the numerics 
almost coincide. Thus, for clarity, we do not show the data points in the figure.
The strong $q$ dependence of the radiation damping is a crucial issue when considering 
the propagation across the chain of an initially localized excitation (see Sec.\ \ref{sec:propagation}).

Expression \eqref{eq:radiationsigma} shows that dark plasmonic modes with a 
wavelength smaller than $\sim 2\pi/k_0$ ($q\gtrsim k_0$), i.e., outside of the 
light cone, have a vanishing radiative decay rate. This behavior arises from the 
destructive interference of the electric field associated with domains of 
in-phase LSPs, resulting in subradiant collective modes that do not couple to light.
The results in Eq.\ \eqref{eq:radiationsigma} and in Fig.~\ref{fig:rad} also show that 
most of the collective plasmons with a wavelength larger than $\sim 2\pi/k_0$ 
($q\lesssim k_0$) are superradiant, with radiative decay rates that exceed that of a 
single nanoparticle $\gamma_0^\mathrm{r}$. 

As illustrated in the inset in Fig.\ \ref{fig:rad}, 
the expression \eqref{eq:radiationsigma} shows that the radiative linewidth 
$\gamma_q^{\sigma, \mathrm{r}}$, scaled with $\gamma_0^\mathrm{r}/k_0d$, is 
almost a universal function of $q/k_0$. In the limit of uncoupled nanoparticles 
($\Omega\rightarrow0$), Eq.\ \eqref{eq:radiationsigma} reduces to 
\begin{equation}
\label{eq:radiationsigma_uncoupled}
 \gamma_{q}^{\sigma,\mathrm{r}}\simeq
 \frac{3\pi|\eta_\sigma|\gamma_0^\mathrm{r}}{4k_0 d}
 \left[1+\mathrm{sgn}\{\eta_\sigma\}\left(\frac{q}{k_0}\right)^2\right]
\Theta\left(k_0-q\right),
\end{equation}
and it is easy to show that $\int\mathrm{d}q\;
\gamma_q^{\sigma, \mathrm{r}}=\pi\gamma_0^\mathrm{r}/d$ 
for both, the transverse and the longitudinal mode. Equation \eqref{eq:radiationsigma_uncoupled} 
demonstrates that the 
radiative linewidth $\gamma_{q}^{\sigma,\mathrm{r}}$ of a chain of noninteracting nanoparticles is significantly different from that of a single nanoparticle $\gamma_0^\mathrm{r}$. This is due to the interference 
effects between the dipolar LSPs in the far field. 
 
The behavior of the radiative decay of the transverse and longitudinal plasmonic 
modes in Fig.~\ref{fig:rad} has been previously addressed by means of sophisticated 
numerical and semi-analytical  calculations, including retardation in the 
interaction between the nanoparticles 
\cite{weber_PRB04,citrin_NL04,citrin_OL06,koend_PRB06,markel_PRB07,petrov_PRA15}. 
Our transparent analytical result \eqref{eq:radiationsigma} shows that a quasistatic 
description of the interparticle interactions is sufficient to describe, at least 
qualitatively, radiative energy losses in metallic nanoparticle chains.

\section{Plasmon propagation along the nanoparticle chain}
\label{sec:propagation}

After having obtained analytical expressions for the nonradiative and radiative lifetimes 
of the collective plasmons in Sec.\ \ref{sec:gamma}, we are now in a position to study
the plasmon propagation along the chain resulting from the irradiation of the first 
nanoparticle by a long-wavelength electric field [cf.\ Eq.\ \eqref{eq:drive}]. 
Toward that end, we introduce the (dimensionless) dipole moment
$\sigma_n=\langle b_n^\sigma+{b_n^{\sigma}}^{\dagger}\rangle$ bared by nanoparticle 
$n$. This quantity  can be 
calculated from its time evolution in momentum space, itself obtained from the master 
equation \eqref{eq:rho} using that 
$\langle\dot{\mathcal{O}}\rangle=\mathrm{Tr}\left\{\dot{\rho}\mathcal{O}\right\}$ 
for any operator $\mathcal{O}$. 
This procedure yields the equation of motion
\begin{equation}
 \label{eq:eom}
\ddot\sigma_q+\gamma_q^\sigma\dot\sigma_q+({\Omega_q^\sigma})^2\sigma_q=\mathcal{A}_q^\sigma f(t), 
\end{equation}
with $\sigma_q=\langle B_q^\sigma+B_q^{\sigma\dagger}\rangle$ 
[cf.\ Eq.\ \eqref{eq:bqs}] and $({\Omega_q^\sigma})^2=
(\tilde{\omega}_q^{\sigma})^2+{(\gamma_q^\sigma/2)}^2$, and 
where the amplitude of the driving force $\mathcal{A}_q^\sigma$ is defined in 
Eq.\ \eqref{eq:A}. In the following, we consider first the case of a continuous drive by 
a monochromatic electric field (Sec.\ \ref{sec:continuous}), and then the case of the irradiation 
of the first nanoparticle by an extremely short ($\delta$-like) laser excitation (Sec.\ \ref{sec:pulse}).

\subsection{Continuous drive by a monochromatic electric field}
\label{sec:continuous}
We start by considering the case in which the first nanoparticle in the chain is illuminated by a long-wavelength 
monochromatic electric field at the driving frequency $\omega_\mathrm{d}$, 
for which $f(t)=\sin{(\omega_\mathrm{d}t)}$.
The stationary solution of Eq.\ \eqref{eq:eom} then reads
\begin{equation}
\label{eq:sigma_q}
\sigma_q=\mathcal{S}_q^\sigma\sin{(\omega_\mathrm{d}t)}+\mathcal{C}_q^\sigma\cos{(\omega_\mathrm{d}t)},
\end{equation}
with 
\begin{subequations}
\label{eq:coefficients}
\begin{equation}
\mathcal{S}_q^\sigma=\mathcal{A}_q^\sigma
\frac{{\Omega_q^\sigma}^2-\omega_\mathrm{d}^2}
{\left(\omega_\mathrm{d}^2-{\Omega_q^\sigma}^2\right)^2+\left(\gamma_q^\sigma\omega_\mathrm{d}\right)^2},
\end{equation}
and
\begin{equation}
\mathcal{C}_q^\sigma=\mathcal{A}_q^\sigma
\frac{-\gamma_q^\sigma\omega_\mathrm{d}}
{\left(\omega_\mathrm{d}^2-{\Omega_q^\sigma}^2\right)^2+\left(\gamma_q^\sigma\omega_\mathrm{d}\right)^2}.
\end{equation}
\end{subequations}
While the time-averaged dipole moment $\overline{\sigma}_n=0$ due to the sinusoidal time dependence in Eq.\  
\eqref{eq:sigma_q} (the bar denotes time averaging), the root-mean-square dipole moment 
$\sqrt{\Delta\sigma_n^2}=\sqrt{\overline{\sigma_n^2}}$ is nonvanishing and 
reads 
\begin{equation}
\label{eq:sigma_n}
\sqrt{\Delta\sigma_n^2}=\frac{1}{\sqrt{\mathcal{N}+1}}
\sqrt{{\left(\mathcal{\tilde{S}}_n^\sigma\right)^2
+\left(\mathcal{\tilde{C}}_n^\sigma\right)^2}
},
\end{equation}
with 
\begin{subequations}
\label{eq:coeff_n}
\begin{equation}
\mathcal{\tilde{S}}_n^\sigma=\sum_q\frac{\sin{(nqd)}}
{\sqrt{\omega_q^\sigma/\omega_0}}\,\mathcal{S}_q^\sigma
\end{equation}
and
\begin{equation}
\mathcal{\tilde{C}}_n^\sigma=\sum_q\frac{\sin{(nqd)}}
{\sqrt{\omega_q^\sigma/\omega_0}}\,\mathcal{C}_q^\sigma.
\end{equation}
\end{subequations}

\subsubsection{Crossover between exponential and algebraic decay of the plasmon excitation along the chain}

\begin{figure*}[tbh]
 \includegraphics[width=1.4\columnwidth]{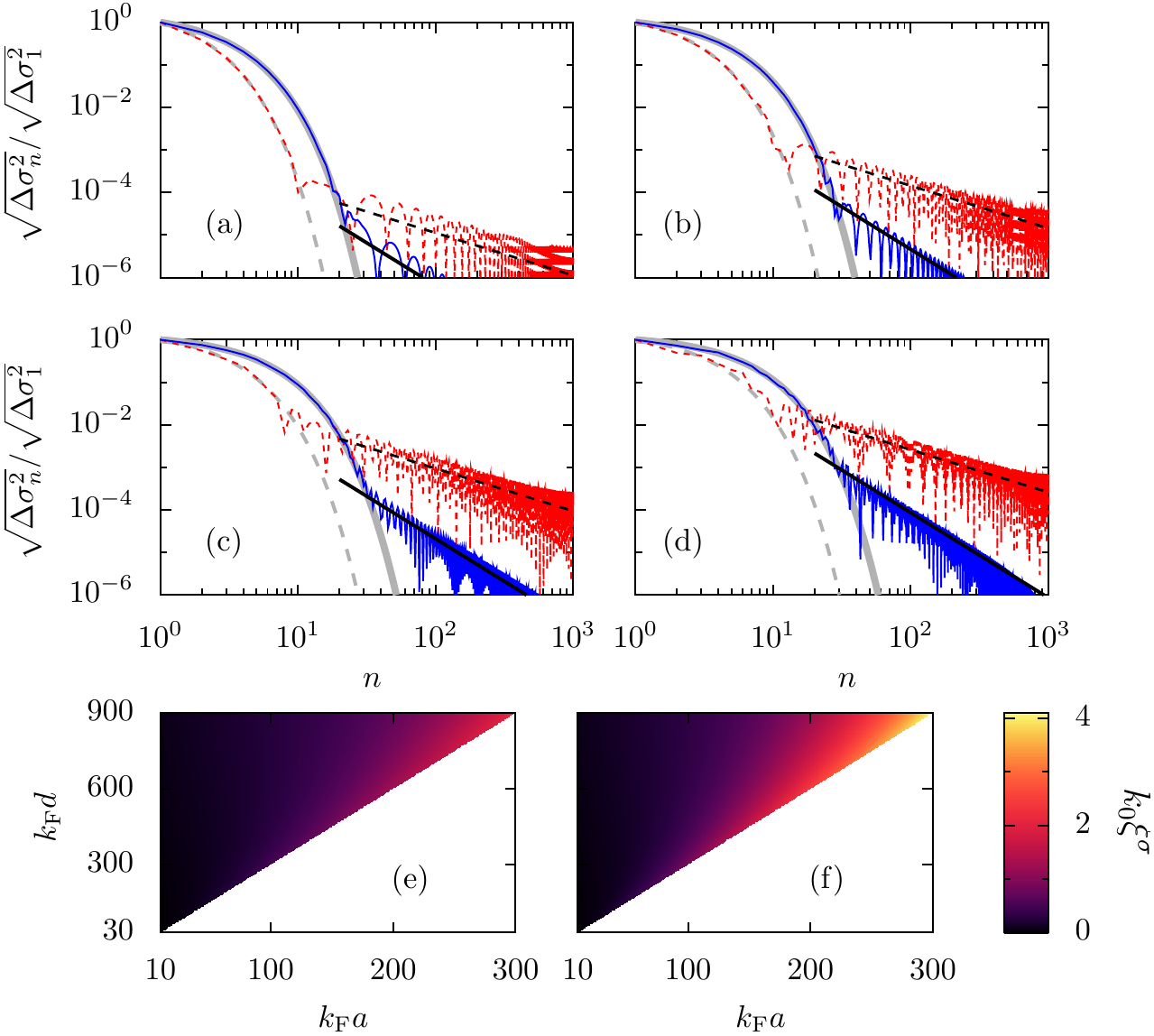}
\caption{\label{fig:propagation}  
(a)--(d) Normalized time-averaged root mean square dipole moment on nanoparticle $n$ 
resulting from a monochromatic excitation at frequency 
$\omega_\mathrm{d}=\omega_0$ 
of the first nanoparticle in a chain with $\mathcal{N}=1000$ and 
interparticle distance $d=3a$. 
Dashed red and solid blue lines correspond, respectively, to a numerical evaluation of Eq.\ \eqref{eq:sigma_n} for the transverse and 
longitudinal modes, including Ohmic losses, Landau damping and radiation damping. 
The thick dashed and solid gray lines corresponding to the analytical result 
\eqref{eq:sigma_n_analytical} include only the nonradiative losses, i.e., Ohmic and Landau damping. 
The nanoparticle sizes are 
(a) $k_\mathrm{F}a=50$, 
(b) $k_\mathrm{F}a=100$, 
(c) $k_\mathrm{F}a=200$, and 
(d) $k_\mathrm{F}a=300$. 
The thick dashed and solid black lines are algebraic
fits for the transverse and longitudinal modes, respectively (see text). 
(e) and (f) Propagation length $\xi^\sigma$ from Eq.\ \eqref{eq:xi_analytical} for the (e) transverse and (f) 
longitudinal modes as a 
function of $a$ and $d$.
The parameters in the figure are
$\gamma^\mathrm{O}/\omega_0=0.027$, 
$\hbar\omega_0/\mathrm{E}_\mathrm{F}=0.47$, and 
$\omega_\mathrm{0}/ck_\mathrm{F}=1.1\times10^{-3}$, corresponding to a chain of Ag nanoparticles.
The frequency shifts \eqref{eq:delqL} and \eqref{eq:delqr} have been neglected.}
\end{figure*}

In Figs.\ \ref{fig:propagation}(a)--(d) we present numerical results for the 
average stationary dipole moment on nanoparticle $n$ 
[cf.\ Eq.\ \eqref{eq:sigma_n}] for the transverse (red dashed lines) and 
longitudinal modes (blue solid lines) 
along a chain composed of $\mathcal{N}=1000$ 
Ag nanoparticles, where the first nanoparticle of the chain is driven at the LSP resonance frequency of the individual 
nanoparticles, $\omega_\mathrm{d}=\omega_0$. 
Panels (a) to (d) in Fig.\ \ref{fig:propagation} correspond 
to nanoparticle radii $k_\mathrm{F}a=50$, $100$, $200$ and $300$, 
respectively, keeping the interparticle distance fixed to $d=3a$.
Here, $k_\mathrm{F}$ denotes the Fermi wavevector.

We observe two different regimes for the decay of the dipole 
moment along the chain:
Over the first few nanoparticles of the chain, the decay of the excitation is purely
exponential,
\begin{equation}
\label{eq:exp}
\frac{\sqrt{\Delta\sigma^2_n}}{\sqrt{\Delta\sigma^2_1}}=\mathrm{e}^{-(n-1)d/\xi^\sigma}, 
\end{equation}
with $\xi^\sigma$ the propagation length for the polarization $\sigma$. 
Remarkably, 
such an exponential decay is exclusively due to the \textit{nonradiative} decay 
mechanisms of the collective plasmons, i.e., Ohmic losses and Landau damping. 
Indeed, the numerical evaluation of Eq.\ \eqref{eq:sigma_n} without the radiation damping \eqref{eq:radiationsigma}
reproduces almost perfectly the exponential decay of $\sqrt{\Delta\sigma^2_n}$ shown in Figs.\ \ref{fig:propagation}(a)--(d).

For longer distances along the chain, the decay of the excitation is algebraic [see the thick solid and 
dashed black lines in Figs.\ \ref{fig:propagation}(a)--(d), which correspond to the fit
$\sqrt{\Delta\sigma^2_n}\sim 1/n^{\zeta^\sigma}$]. 
This algebraic decay results solely from the radiation damping \eqref{eq:radiationsigma} and its behavior 
as a function of 
momentum. The latter for the transverse modes is discontinuous 
(for $\mathcal{N}\gg1$) at $q\simeq k_0$ [see Fig.\ \ref{fig:rad}(a)], 
yielding $\zeta^{x,y}\simeq1$ [see the thick dashed black lines in 
Figs.\ \ref{fig:propagation}(a)--(d)], while for the longitudinal mode [see 
Fig.\ \ref{fig:rad}(b)], the monotonic decaying behavior of the radiative 
damping rate for $q\lesssim k_0$ yields $\zeta^z\simeq2$.
The algebraic behavior of the plasmon decay along the chain becomes more 
predominant for increasing nanoparticle size and interparticle distance, 
at the constant ratio $d=3a$ used in Figs.\ \ref{fig:propagation}(a)--(d).
This is a result of the increasing influence of the radiation damping on 
the overall collective plasmon linewidth for increasing nanoparticle sizes.

\subsubsection{Propagation length}
\label{sec:analytical}
Within the perspective of energy transfer, the initial exponential regime witnessed in Figs.\ \ref{fig:propagation}(a)--(d) is the determinant one. 
Therefore, it is useful to search for the maximization of the propagation length $\xi^\sigma$ defined in Eq.\ \eqref{eq:exp}
within the parameter range of the present model. Below we provide an analytical calculation of the root-mean-square dipole moment \eqref{eq:sigma_n} in the weakly-coupled nanoparticle regime, 
and subsequently we deduce the propagation length $\xi^\sigma$ as a function of the parameters of our model.

Since the exponential decay of the plasmon excitation is of nonradiative 
origin, we neglect in what follows the radiation damping \eqref{eq:radiationsigma}. Moreover, we neglect the frequency 
shifts \eqref{eq:delqL} and \eqref{eq:delqr} as these represent a very small correction to the collective mode resonance frequencies \cite{prb2006}. 
To linear order in the coupling \eqref{eq:Omega} and to quadratic order in $\gamma_q^\sigma/\omega_0$, using Eqs.\ \eqref{eq:plasmon_spectrum} and \eqref{eq:landau}, 
the coefficients \eqref{eq:coefficients} read for $\omega_\mathrm{d}=\omega_0$
\begin{subequations}
\label{eq:coefficients_approx}
\begin{equation}
\mathcal{S}_q^\sigma\simeq\frac{\mathcal{A}_q^\sigma}{4\omega_0^2}
\frac{\eta_\sigma\Omega\omega_0\cos{(qd)}+\left(\gamma^\mathrm{nr}/4\right)^2}
{\left[\eta_\sigma\Omega\cos{(qd)}\right]^2+\left(\gamma^\mathrm{nr}/4\right)^2}
\end{equation}
and
\begin{equation}
\mathcal{C}_q^\sigma\simeq-\frac{\mathcal{A}_q^\sigma}{8\omega_0^2}
\frac{\gamma^\mathrm{nr}\omega_0/2+\eta_\sigma\gamma_0^\mathrm{L}\Omega G(\hbar\omega_0/E_\mathrm{F})\cos{(qd)}}
{\left[\eta_\sigma\Omega\cos{(qd)}\right]^2+\left(\gamma^\mathrm{nr}/4\right)^2},
\end{equation}
\end{subequations}
where $\gamma^\mathrm{nr}=\gamma^\mathrm{O}+\gamma_0^\mathrm{L}$ is the nonradiative part of the damping rate corresponding to a single nanoparticle. 
In the large chain limit ($\mathcal{N}\gg1$), we replace the summation over plasmon momenta in Eq.\ \eqref{eq:coeff_n} by 
an integral, and we arrive, using Eq.\ \eqref{eq:coefficients_approx}, at
\begin{subequations}
\label{eq:coefficients_calculated}
\begin{align}
\mathcal{\tilde{S}}_n^\sigma\simeq&
-\frac{\sqrt{2(\mathcal{N}+1)}}{2\pi\eta_\sigma}\, \hat\sigma\cdot\hat\epsilon\,\frac{\Omega_\mathrm{R}}{\Omega}
\left[
\frac{(\gamma^\mathrm{nr}/4)^2}{\eta_\sigma\Omega\omega_0}
\mathcal{I}_n\left(\frac{\gamma^\mathrm{nr}}{4|\eta_\sigma|\Omega}\right)
\right.
\nonumber\\
&+\left.\mathcal{J}_n\left(\frac{\gamma^\mathrm{nr}}{4|\eta_\sigma|\Omega}\right)
\right]
\end{align}
and
\begin{align}
\mathcal{\tilde{C}}_n^\sigma\simeq&\;
\frac{\sqrt{2(\mathcal{N}+1)}}{4\pi\eta_\sigma}\,\hat\sigma\cdot\hat\epsilon\,\frac{\Omega_\mathrm{R}}{\Omega}
\left[
\frac{\gamma^\mathrm{nr}}{2\eta_\sigma\Omega}
\mathcal{I}_n\left(\frac{\gamma^\mathrm{nr}}{4|\eta_\sigma|\Omega}\right)
\right.
\nonumber\\
&+\left.
G\left(\frac{\hbar\omega_0}{E_\mathrm{F}}\right)\frac{\gamma_0^\mathrm{L}}{\omega_0}
\mathcal{J}_n\left(\frac{\gamma^\mathrm{nr}}{4|\eta_\sigma|\Omega}\right)
\right], 
\end{align}
\end{subequations}
where the Rabi frequency $\Omega_\mathrm{R}$ is introduced in Eq.\ \eqref{eq:Rabi}.
The integrals $\mathcal{I}_n$ and $\mathcal{J}_n$ are defined in Eqs.\ \eqref{eq:I_n} and \eqref{eq:J_n}, respectively, and they are evaluated in Appendix \ref{sec:integrals}. 

With Eqs.\ \eqref{eq:I_n_final} and \eqref{eq:J_n_final} and to leading order in 
$\Omega/\omega_0$ and $\gamma_q^\sigma/\omega_0$, Eq.\ \eqref{eq:sigma_n} finally reads
\begin{equation}
\label{eq:sigma_n_analytical} 
\sqrt{\Delta\sigma_n^2}\simeq
\frac{|\hat\sigma\cdot\hat\epsilon|}{\sqrt{2}|\eta_\sigma|}\frac{\Omega_\mathrm{R}}{\Omega}
\left[
\sqrt{1+\left(\frac{\gamma^\mathrm{nr}}{4|\eta_\sigma|\Omega}\right)^2}-\frac{\gamma^\mathrm{nr}}{4|\eta_\sigma|\Omega}
\right]^n.
\end{equation}
The decay of the plasmon excitation then follows the exponential behavior \eqref{eq:exp}, with a decay length \begin{equation}
\label{eq:xi_analytical}
\xi^\sigma=\frac{d}{\mathrm{arcsinh}(\gamma^\mathrm{nr}/4|\eta_\sigma|\Omega)}.
\end{equation}
The latter, once scaled with the interparticle distance $d$, is a monotonically decreasing function of the unique parameter $\gamma^\mathrm{nr}/\Omega$. For weak dissipation and/or strong coupling ($\gamma^\mathrm{nr}\ll\Omega$), it behaves as $\xi^\sigma/d\simeq4|\eta_\sigma|\Omega/\gamma^\mathrm{nr}$, while in the opposite regime $\gamma^\mathrm{nr}\gg\Omega$, $\xi^\sigma/d\simeq[\ln{(\gamma^\mathrm{nr}/4|\eta_\sigma|\Omega)}+\ln{2}]^{-1}$.

We show in Figs.\ \ref{fig:propagation}(a)--(d) our analytical result \eqref{eq:sigma_n_analytical} for the transverse and longitudinal modes by thick solid and dashed gray lines, respectively. As can be seen from the figure, the agreement between Eq.\ \eqref{eq:sigma_n_analytical} and the exponential part of the plasmon decay as obtained from the numerics is excellent, confirming that such an exponential behavior is solely of nonradiative origin. 

The propagation length \eqref{eq:xi_analytical} is plotted in Fig.\ \ref{fig:propagation}(e) for the transverse modes 
and in Fig.\ \ref{fig:propagation}(f) for the longitudinal one
for chains of Ag nanoparticles. The propagation length $\xi^\sigma$ \ is measured in units of $k_0^{-1}=\unit[76]{nm}$, while $a$ and $d$ 
are measured in units of $k_\mathrm{F}^{-1}=\unit[0.83]{\AA}$. 
In these two figures, we only show data points for $d\geqslant3a$, as our model of point dipoles interacting through a quasistatic interaction is 
not valid for smaller interparticle distances \cite{park_PRB04}. 

As can be seen from panels (e) and (f) in Fig.~\ref{fig:propagation} and inferred from Eq.\ \eqref{eq:xi_analytical}, 
the smaller the interparticle distance $d$ 
and the larger the nanoparticle radii $a$, i.e., the larger the coupling constant $\Omega$ defined in Eq.\ \eqref{eq:Omega}, the larger is the propagation length $\xi^\sigma$. 
For a fixed $d$, the maximum $\xi^\sigma$ is attained for $d/a=3$, that is, at the limit of validity of the near-field approximation adopted in this work. It is then 
expected that the optimal propagation lengths occur for $d/a<3$. 

An important conclusion that can be extracted from Figs.\ \ref{fig:propagation}(e) and \ref{fig:propagation}(f) and from Eq.\ \eqref{eq:xi_analytical} is that the longitudinal mode 
generally propagates for a longer distance than the transverse one, thus confirming previous numerical studies \cite{Quin98_OL, Brog00_PRB, meier_PRB03}
in the framework of a well-defined criterion. This is due to the fact that the LSPs have an effective interaction strength in Eq.\ \eqref{eq:Hpl} that is twice as large in the longitudinal
case ($|\eta_z|=2$) as in the transverse case ($|\eta_{x,y}|=1$).

\subsection{Short laser pulse}
\label{sec:pulse}
We now consider an alternative situation of experimental relevance, where the first nanoparticle in the chain 
is irradiated by a very short laser pulse. In such a case, $f(t)=\delta(\omega_0t)$ and
the solution of Eq.\ \eqref{eq:eom} can be readily obtained, yielding 
the (dimensionless) dipole moment on nanoparticle $n$, 
\begin{align}
\label{eq:sigma_n_pulse}
\sigma_n(t)=&\;\sqrt{\frac{2}{(\mathcal{N}+1)\omega_0}}\Theta(t)\sum_q
\frac{\mathcal{A}_q^\sigma}{(\omega_q^\sigma)^{3/2}}\sin{(nqd)}
\nonumber\\
&\times
\mathrm{e}^{-\gamma_q^\sigma t/2}\sin{(\omega_q^\sigma t)}.
\end{align}

\begin{figure*}[tbh]
 \includegraphics[width=\linewidth]{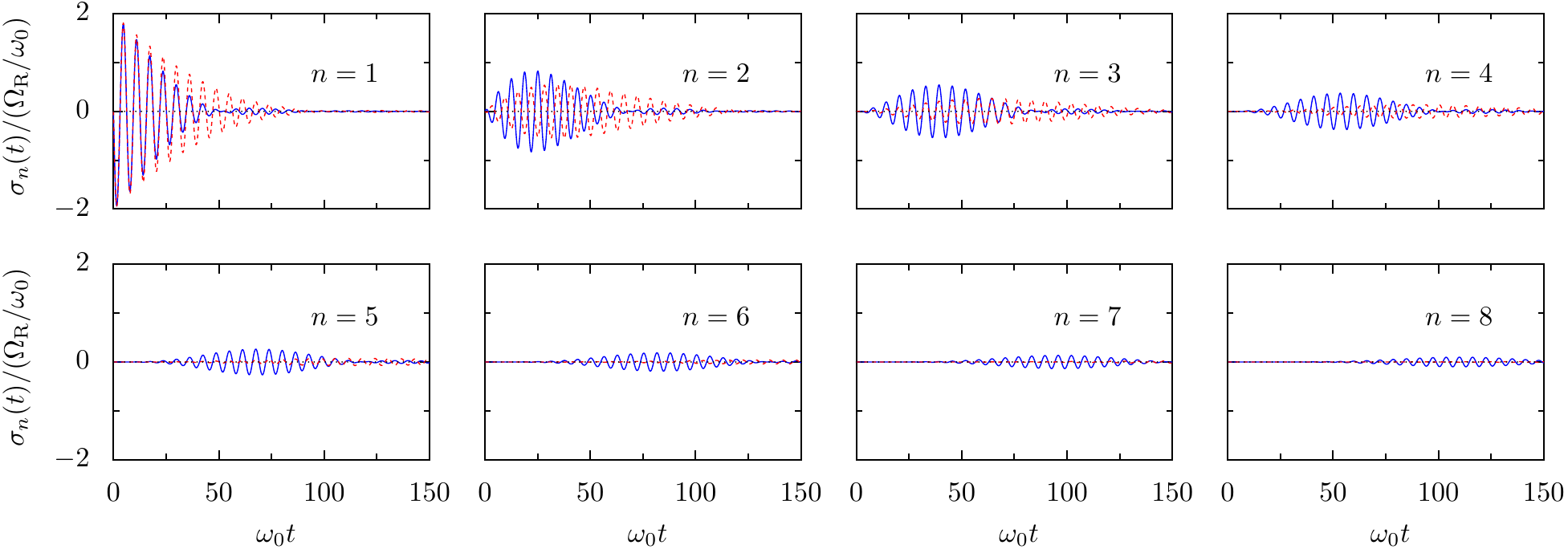}
\caption{\label{fig:pulse}  
Dipole moment on nanoparticle $n$ as a function of time resulting from the excitation of the first nanoparticle by a short pulse [cf.\ Eq.\ \eqref{eq:sigma_finalresult}]. 
The transverse and longitudinal modes are represented by red dashed and blue solid lines, respectively.  
The parameters used in the figure correspond to an infinite chain of Ag nanoparticles with radius $a=200\ k_\mathrm{F}^{-1}=\unit[16.6]{nm}$ separated 
by an interparticle distance $d=3a$.}
\end{figure*}

We have checked by a numerical 
evaluation of Eq.\ \eqref{eq:sigma_n_pulse} (not shown) that the radiation damping weakly affects the decay of the plasmon 
excitation along the chain for short distances (below ca.\ 10 nanoparticles), as is the case for the continuous drive by a monochromatic field (see Sec.\ \ref{sec:continuous}). 
Along the same lines as in the previous section, in the following we thus evaluate Eq.\ \eqref{eq:sigma_n_pulse} analytically by disregarding the radiation damping \eqref{eq:radiationsigma}. We further ignore the frequency shifts \eqref{eq:delqL} and \eqref{eq:delqr}. In the 
large chain limit and working up to leading order in $\Omega/\omega_0\ll 1$, we then obtain 
\begin{align}
\label{eq:sigma_n_pulse_2}
\sigma_n(t)\simeq&-\frac{4}{\pi}\ \hat\sigma\cdot\hat\epsilon\ \frac{\Omega_\mathrm{R}}{\omega_0}\
\Theta(t)\ \mathrm{e}^{-\gamma^\mathrm{nr}t/2}
\nonumber\\
&\times
\left[
\mathcal{K}_n\left(2\eta_\sigma\Omega t, \eta_\sigma\gamma_0^\mathrm{L}G(\hbar\omega_0/E_\mathrm{F})\Omega t/\omega_0\right) \sin{(\omega_0t)}
\right.
\nonumber\\
&\left.+\ \  \mathcal{L}_n\left(2\eta_\sigma\Omega t, \eta_\sigma\gamma_0^\mathrm{L}G(\hbar\omega_0/E_\mathrm{F})\Omega t/\omega_0\right) \cos{(\omega_0t)}
\right].
\end{align}
The integrals $\mathcal{K}_n$ and $\mathcal{L}_n$ are defined in Eqs.\ \eqref{eq:K_def} and \eqref{eq:L_def} and 
are evaluated in Appendix \ref{sec:details}. Together with Eqs.\ \eqref{eq:K_result} and \eqref{eq:L_result}, we then obtain to leading order in $\Omega/\omega_0$ and in $\gamma_q^\sigma/\omega_0$ the result
\begin{align}
\label{eq:sigma_finalresult}
\sigma_n(t)=&\;\frac{2\Omega_\mathrm{R}}{\omega_0}\ \hat\sigma\cdot\hat\epsilon\ \Theta(t)
\nonumber\\
&\times
\frac{\mathrm{e}^{-\gamma^\mathrm{nr} t/2}\cos{(\omega_0 t+n\pi/2)}\ nJ_n(2\eta_\sigma\Omega t)}{\eta_\sigma\Omega t}
\end{align}
for the dipole moment on nanoparticle $n$ resulting from a pulsed excitation on the first nanoparticle in the chain. In the previous expression, 
$J_n(z)$ denotes the Bessel function of the first kind.

In Fig.\ \ref{fig:pulse} we plot the dipole moment \eqref{eq:sigma_finalresult} on nanoparticle $n=1$ to $8$ as a function of time for the 
transverse (red dashed lines) and longitudinal modes (blue solid lines). The parameters used in the figure correspond to 
the case of an infinite chain of Ag nanoparticles with radius $a=200\ k_\mathrm{F}^{-1}=\unit[16.6]{nm}$ and interparticle distance $d=3a$. 
As can be seen in the figure, the initial excitation propagates for at least 
$n=8$ nanoparticles in the case of the longitudinal mode, corresponding to a distance of about $\unit[400]{nm}$. 
It is clear from Fig.\ \ref{fig:pulse} for $n=1$ to $4$ that the transverse 
mode (cf.\ the red dashed lines in the figure) has a longer lifetime than the longitudinal one (blue solid lines). 
Such a longer lifetime  is associated with lower propagation efficiency.
Hence, the longitudinal mode propagates for 
longer distances than the transverse mode. For instance on the 8$^\textrm{th}$ nanoparticle, the longitudinal mode is still active (at the level of a few percent of the initial excitation) while the 
transverse mode is totally suppressed. One may conclude from Fig.\ \ref{fig:pulse} that, although the signal is strongly damped, it 
may still be detectable and therefore may be useful in the prospect of information transfer based on nanoscale plasmonic metamaterials.

\section{Conclusions}
\label{sec:ccl}
We have considered collective plasmonic excitations in finite and infinite chains of spherical metallic nanoparticles, and in particular their damping. 
Our open quantum system approach has enabled us to quantify the two most 
important size-dependent damping mechanisms that lead to the decay of the 
plasmonic excitations along the chain: Landau damping due to the coupling 
to internal electronic degrees of freedom and radiation damping due to the
coupling to the surrounding electromagnetic field modes. 
We have derived and presented a universal analytical formula for the 
nonradiative Landau damping decay rate of coupled plasmonic modes that 
is valid for an arbitrary chain length. We have shown that Landau damping is 
predominant for small nanoparticles as it scales as 
the inverse of their diameter. Moreover, unlike radiative losses, Landau 
damping is nonzero for all plasmon modes of any wavelength. 

We have calculated the radiation damping decay rate of the coupled modes 
and obtained an analytical expression for the infinite chain limit. 
We have performed numerical calculations for finite chains, thereby extending 
the investigated parameter regime and confirming the analytically predicted 
behavior for infinite chains. 
Our transparent analytical results, obtained within a quasistatic 
approximation, are also consistent with existing numerical results which 
include retardation effects. 
Therefore, we can conclude that retardation effects do not play a crucial role
and are rather unimportant for explaining, at least qualitatively, 
radiation losses in coupled plasmonic nanostructures.  

Using the reduced density matrix formalism, we have investigated 
the decay of the plasmonic excitation along the chain when a
long-wavelength laser field illuminates the first nanoparticle of the chain. In the case of a
continuous drive by a monochromatic field, 
we have numerically demonstrated that there are two distinct regimes for the 
decay of the plasmonic modes along the chain. 
For short distances (typically of the order of $10$ nanoparticles), the decay of the 
plasmonic excitation exhibits an exponential behavior along the chain. 
Importantly, we have shown that 
such an exponential decay is due solely to the nonradiative damping mechanisms
(i.e., Ohmic losses and Landau damping), and that it is not influenced by radiation damping.
For longer distances, 
the decay becomes algebraic, with a polarization-dependent power law. 
Such an algebraic decay is exclusively due to the behavior of the radiation 
damping decay rate as a function of momentum. 
This regime switching is of foremost importance for characterizing the alteration of an excitation along 
the chain, and it must be taken into account when comparing the different predictions and measurements of the 
decay lengths. 
Similar conclusions can be drawn from the case of a pulsed excitation on the first nanoparticle. 

We have provided transparent and simple analytical 
expressions for the exponentially-decaying plasmon excitation profile along the chain and its associated 
plasmon propagation length, which is larger for the longitudinal mode 
than for the two transverse modes. The largest propagation lengths were found at the limit of validity of the 
present model. Therefore, it is desirable to develop alternative models in order to extend the parameter range 
explored in this work. In addition, 
it would be interesting to extend 
the theory of the decay of coupled plasmonic modes presented in this paper, 
which captures the essential physics of the problem, 
to other one- and two-dimensional arrays of metallic nanoparticles, 
presenting, e.g., interesting topological features, such as the honeycomb array 
baring chiral bosonic Dirac plasmons \cite{Weick_PRL, Sturges_2DMat}.

\begin{acknowledgments}
We are grateful to P.\ W.\ Brouwer, E.\ Mariani, 
S.\ Mukamel and F.\ Vall\'{e}e 
for useful discussions. We acknowledge financial support from the CNRS through 
the PICS program (Contract No.\ 6384 APAG) and from the ANR under Grant No.\ 
ANR-14-CE26-0005 Q-MetaMat.
\end{acknowledgments}

\appendix
\section{The need to go beyond the rotating-wave approximation for the plasmonic 
Hamiltonian}
\label{sec:RWA}
In this appendix, we briefly comment on the rotating-wave approximation (RWA) 
for the plasmonic Hamiltonian \eqref{eq:Hpl}.
We show that, although it gives correct results for the dispersion relation 
to first order in the small parameter
$\Omega/\omega_0\ll1$ defined in Eq.\ \eqref{eq:Omega}, it misses corrections of the same order in 
$\Omega/\omega_0$ to the eigenstates. These are, however, important for 
state-dependent quantities such as the nonradiative and radiative 
collective plasmon decay rates evaluated in Sec.\ \ref{sec:gamma} in the 
general case and in Appendix \ref{sec:dimer} for the special case of a heterogeneous 
nanoparticle dimer. 

The RWA counterpart of the Hamiltonian \eqref{eq:Hpl} reads 
\begin{align}
 H_\mathrm{pl}^\mathrm{RWA}=&\;
 \hbar\omega_0\sum_{n=1}^\mathcal{N}\sum_{\sigma=x,y,z} 
 {b_{n}^{\sigma}}^{\dagger} b_{n}^{\sigma}
 \nonumber\\
 &+
 \hbar\Omega\sum_{n=1}^{\mathcal{N}-1}\sum_{\sigma=x,y,z} 
\eta_\sigma
\left(b_{n}^{\sigma} {b_{n+1}^{\sigma}}^{\dagger}+b_{n+1}^{\sigma} {b_{n}^{\sigma}}^{\dagger}\right)
\end{align}
and it is easily diagonalized using the sine transform \eqref{eq:sinetransform} 
to yield
\begin{equation}
H_\mathrm{pl}^\mathrm{RWA}=\sum_{q\sigma}
\hbar\omega_q^{\sigma,\mathrm{RWA}}{b_q^{\sigma}}^{\dagger}b_q^\sigma.
\end{equation}
Within the RWA, the plasmon dispersion reads 
\begin{equation}
\omega_q^{\sigma,\mathrm{RWA}}=\omega_0+2\eta_\sigma\Omega\cos{(qd)},
\end{equation} 
coinciding with the exact spectrum \eqref{eq:plasmon_spectrum} to first 
order in $\Omega/\omega_0\ll1$. However, the RWA leads to values of the
Bogoliubov coefficients $\cosh{\theta_q^\sigma}$ and $\sinh\theta_q^\sigma$ 
[cf.\ Eq.\ \eqref{eq:Bogoliubov}] that are $1$ and $0$, respectively, while these coefficients read, to first order in $\Omega/\omega_0$, as $\cosh\theta_q^\sigma\simeq1$ and 
$\sinh\theta_q^\sigma\simeq\eta_\sigma(\Omega/\omega_0)\cos{(qd)}$. Hence, the RWA misses the latter correction to the plasmon eigenstates, and therefore it gives incorrect results for state-dependent quantities.

\section{The case of a heterogeneous nanoparticle dimer}
\label{sec:dimer}
The technical ideas presented in this work can be easily tested on the special 
case of a nanoparticle dimer ($\mathcal{N}=2$), even in the 
heterogeneous case of different nanoparticle sizes and/or made of different 
materials. The analytical
results obtained within the present approach can be 
checked against previous 
developments \cite{my}. For completeness, we adapt the formulation of the main text to 
the specific case at hand.
The 
plasmonic part of the Hamiltonian \eqref{eq:H} now reads 
\begin{align}
 \label{eq:aHpl}
 H_\mathrm{pl}=&\;\sum_{n=1}^2\sum_{\sigma=x,y,z}
 \hbar\omega_n b^{\sigma\dagger}_n b_n^\sigma
 \nonumber\\
& +\hbar\Omega\sum_{\sigma=x,y,z}
\eta_\sigma\left( b_1^\sigma+{b_1^{\sigma}}^{\dagger} \right)
\left( b_2^\sigma+{b_2^{\sigma}}^{\dagger} \right),
\end{align}
where $\omega_n$ is the LSP resonance frequency in the 
$n$\textsuperscript{th} nanoparticle, and it corresponds for simple metals 
and neglecting the spill-out effect to the 
Mie frequency $(N_ne^2/m_\mathrm{e}a_n^3)^{1/2}$, with $N_n$ and $a_n$ the 
electron number and the radius of nanoparticle $n$, respectively.
The coupling frequency reads 
\begin{equation}
\Omega=\frac{\sqrt{\omega_1\omega_2}}{2} \left(\frac{\sqrt{a_1a_2}}{d}\right)^3.
\end{equation} 

The Hamiltonian \eqref{eq:aHpl} is diagonalized to 
\begin{equation}
 \label{eq:aHplq}
 H_\mathrm{pl}=\sum_\sigma\left(\hbar\omega_+^\sigma {B_+^{\sigma}}^{\dagger}B_+^\sigma
 +\hbar\omega_-^\sigma {B_-^{\sigma}}^{\dagger}B_-^\sigma\right)
\end{equation}
by a Bogoliubov transformation \cite{tsall78_JMP}
\begin{equation}
\label{eq:bog_dimer}
B_\pm^\sigma=\sum_{n=1}^2\left(u_{n\pm}^\sigma b_n^\sigma+\bar{u}_{n\pm}^\sigma b_n^{\sigma\dagger}\right),
\end{equation}
where the eigenfrequencies of the high- and low-energy eigenmodes read
\begin{align} 
\label{eq:awx}
 \omega_\pm^{\sigma}=\sqrt{\frac{{\omega}_1^2+{\omega}_2^2}{2}
\pm\sqrt{4\eta_\sigma^2\Omega^2{\omega}_1 {\omega}_2
+\left(\frac{{\omega}_1^2-{\omega}_2^2}{2}\right)^2}},
\end{align}
respectively. 
The high-energy (low-energy) mode for the transverse polarization 
$\sigma=x=y$ corresponds to a bright (dark) mode. Vice versa, 
the high-energy (low-energy) mode for the longitudinal $\sigma=z$ polarization 
corresponds to a dark (bright) mode.
The coefficients entering Eq.\ \eqref{eq:bog_dimer} read
\begin{subequations}
\begin{align}
 u^\sigma_{n,\pm}=[\pm \mathrm{sgn}\{\eta_\sigma\}]^{n-1}
\frac{\omega_\pm^{\sigma}+{\omega}_n}{2\sqrt{{\omega}_n\omega_\pm^{\sigma}}}
\sqrt{\frac{{\omega_\pm^{\sigma}}^2-{\omega}_{\hat n}^2}{2{\omega_\pm^{\sigma}}^2-{\omega}_1^2-{\omega}_2^2}},\\
 \bar u^\sigma_{n,\pm}=[\pm \mathrm{sgn}\{\eta_\sigma\}]^{n-1}
\frac{\omega_\pm^{\sigma}-{\omega}_n}{2\sqrt{{\omega}_n\omega_\pm^{\sigma}}}
\sqrt{\frac{{\omega_\pm^{\sigma}}^2-{\omega}_{\hat n}^2}{2{\omega_\pm^{\sigma}}^2-{\omega}_1^2-{\omega}_2^2}},
\end{align}
\end{subequations}
where $\hat n=1(2)$ for $n=2(1)$.

The coupling between plasmons and electron-hole pairs reads for a 
heterogeneous dimer as
\begin{align}
 \label{eq:aHpleh}
 H_\mathrm{pl\textrm{-}eh}=&\;\sum_{n=1}^2\sum_{\sigma=x,y,z}\Lambda_n
 \left(b_n^\sigma+{b_n^{\sigma}}^{\dagger}\right)
 \nonumber\\
&\times \sum_{\alpha\beta}\left<n\alpha\right|\hat{\sigma}\cdot \mathbf{r}_n\left|n\beta\right>c_{n\alpha}^\dagger c_{n\beta}^{\phantom{\dagger}},
\end{align}
with $\mathbf{r}_n$ the vector originating from the center of 
nanoparticle $n$. 
The Landau damping of the heterogeneous dimer can then be expressed 
as \cite{my}
\begin{equation}
 \label{eq:alandau}
 \gamma_\pm^{\sigma,\rm L}=\sum_{n=1}^2\frac{3v_{\rm F}^{(n)}}{4a_n}\left(\frac{\omega_n}{\omega_\pm^\sigma}\right)^3
 g\left(\frac{\hbar\omega_\pm^\sigma}{E_{\rm F}^{(n)}}\right)\left(\Delta u_{n\pm}^\sigma\right)^2,
\end{equation}
where $v_{\rm F}^{(n)}$ and $E_{\rm F}^{(n)}$ are the Fermi velocity and 
energy of the $n$\textsuperscript{th} nanoparticle, respectively, and where 
the function $g(\nu)$ is defined in Eq.\ \eqref{eq:g}. In the expression 
above, $\Delta u_{n\pm}^\sigma=u_{n\pm}^\sigma-\bar{u}_{n\pm}^\sigma$. 
For a homogeneous dimer made of two nanoparticles of the same size and  
material, the Landau damping linewidth \eqref{eq:alandau} reduces to 
\begin{equation}
 \gamma_\pm^{\sigma,\mathrm{L}}=\frac{3v_\mathrm{F}}{4a}
 \left(\frac{\omega_0}{\omega_\pm^\sigma}\right)^4
g\left(\frac{\hbar\omega_\pm^\sigma}{E_{\rm F}}\right), 
\end{equation}
corresponding to Eq.\ \eqref{eq:landau} for $\mathcal{N}=2$. 

In Eq.\ \eqref{eq:H}, the coupling between plasmons and photons for a 
heterogeneous dimer reads in the dipolar approximation 
\begin{align}
 \label{eq:aHplph}
 H_\mathrm{pl\textrm{-}ph}=&\;
 \mathrm{i}\hbar\sum_{n=1}^{2}\sum_{\sigma=x,y,z}\sum_{\mathbf{k},\hat{\lambda}_\mathbf{k}} 
 \sqrt{\frac{\pi\omega_n^3 a_n^3  }{{\mathcal{V}\omega_\mathbf{k}}}}
 \hat{\sigma}\cdot \hat{\lambda}_{\mathbf{k}}
 \nonumber\\
 &\times
 \left({b_{n}^{\sigma}}^{\dagger}-b_{n}^{\sigma}\right)
  \left(a_{\mathbf{k}}^{\hat{\lambda}_\mathbf{k}} \,
 \mathrm{e}^{\mathrm{i}\mathbf{k}\cdot \mathbf{d}_n}
 +{a_{\mathbf{k}}^{\hat{\lambda}_\mathbf{k}}}^\dagger\,
 \mathrm{e}^{-\mathrm{i}\mathbf{k}\cdot \mathbf{d}_n} \right).
\end{align}
With the help of the Bogoliubov transformation \eqref{eq:bog_dimer}, 
the Fermi golden rule radiative decay rate hence takes the form
\begin{align}
 \label{eq:agammaR}
 \gamma_\pm^{\sigma,\mathrm{r}}=&\;\frac{2\pi^2}{\mathcal{V}}
 \sum_{\mathbf{k},\hat{\lambda}_\mathbf{k}}
\frac{{|\hat{\sigma}\cdot \hat{\lambda}_\mathbf{k}|}^2}{\omega_\mathbf{k}}
\left|
\sum_{n=1}^2(\omega_n a_n)^{3/2}U_{n\pm}^\sigma\mathrm{e}^{-\mathrm{i}\mathbf{k}\cdot \mathbf{d}_n}
\right|^2 
\nonumber\\
&\times\delta(\omega^{\sigma}_\pm-\omega_\mathbf{k}),
\end{align}
where $U_{n\pm}^\sigma=u_{n\pm}^\sigma+\bar{u}_{n\pm}^\sigma$. 
After summing over photon polarizations, and replacing in the limit 
$\mathcal{V}\rightarrow\infty$ the summation over photon momenta by an 
integral, we arrive at the general result for the radiation damping of a 
heterogeneous metallic nanoparticle dimer, 
\begin{align}
\label{eq:gamma_rad_dim}
\gamma_\pm^{\sigma, \mathrm{r}}=&\;
\frac{2\omega_\pm^\sigma}{3c^3}
\Bigg\{
\sum_{n=1}^2(\omega_n a_n)^3{U_{n\pm}^\sigma}^2
+3\prod_{n=1}^2(\omega_na_n)^{3/2}U_{n\pm}^\sigma
\nonumber\\
&\times
\Bigg[
\left(\frac{\Theta(\eta_\sigma)}{k_\pm^\sigma d}-\frac{\eta_\sigma}{(k_\pm^\sigma d)^3}\right)\sin{(k_\pm^\sigma d)}
\nonumber\\
&+\frac{\eta_\sigma}{(k_\pm^\sigma d)^2}\cos{(k_\pm^\sigma d)}
\Bigg]
\Bigg\}, 
\end{align}
with $k_\pm^\sigma=\omega_\pm^\sigma/c$. 
In the limit $k_\pm^\sigma d\ll 1$, the above expression reduces to 
\begin{equation}
 \label{eq:aradiation}
 \gamma_\pm^{\sigma, \mathrm{r}}=\frac{2{\omega_\pm^\sigma}^3}{3c^3}\left(\sum_{n=1}^2 \sqrt{\omega_na_n^3}\Delta u_{n\pm}^\sigma  \right)^2,
\end{equation}
and we thus recover the result of Ref.\ \cite{my}. 

\begin{figure}[tb]
 \includegraphics[width=\columnwidth]{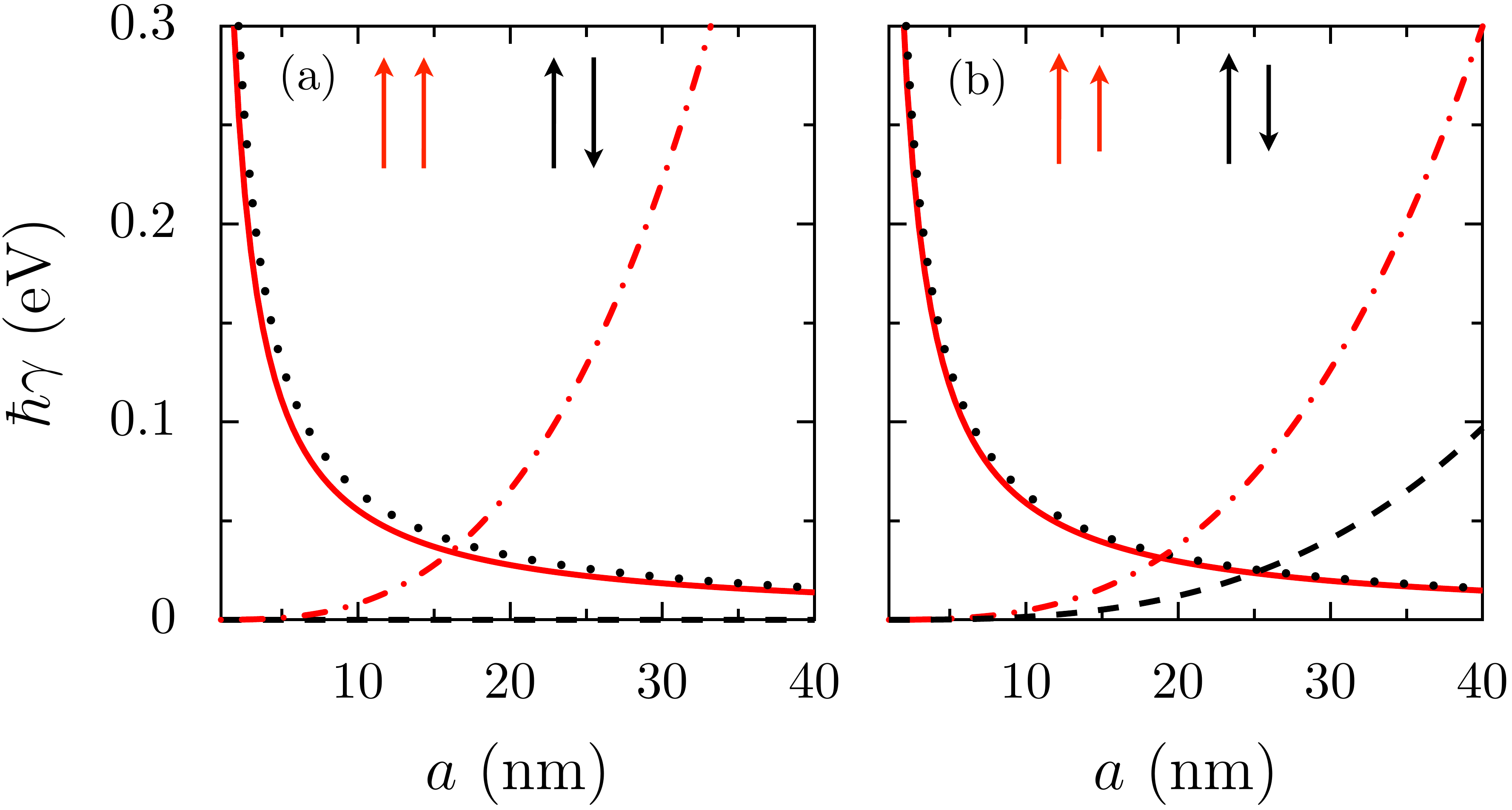}
\caption{\label{fig:gamma_dimer} 
Landau damping (solid and dotted lines) and radiation damping 
(dashed-dotted and dashed lines) decay rates of the transverse mode
as a function of nanoparticle radius $a$ for the bright 
($+$, light gray/red lines) and dark ($-$, black lines) mode of a 
nanoparticle dimer ($\mathcal{N}=2$) with $d=3a$. 
(a) Homogeneous dimer formed by two identical Ag nanoparticles and (b) 
heterogeneous Ag-Au dimer 
embedded in a medium with dielectric constant $\epsilon_\mathrm{m}=4$.}
\end{figure}

We show in Fig.\ \ref{fig:gamma_dimer} the competition between the Landau and 
radiative decay rates of the bright (light gray/red lines) and dark (black 
lines) plasmonic transverse modes as a function of nanoparticle radius $a$ 
(assumed to be the same for both particles) for a homogeneous Ag-Ag [Fig.\ 
\ref{fig:gamma_dimer}(a)] and heterogeneous Ag-Au dimer [Fig.\ 
\ref{fig:gamma_dimer}(b)] with interparticle separation $d=3a$. 
We assume that the dimers are embedded in a medium with dielectric constant 
$\epsilon_\mathrm{m}=4$, corresponding to LSP resonance frequencies 
$\omega_\mathrm{Ag}=\unit[2.6]{eV/\hbar}$ and 
$\omega_\mathrm{Au}=\unit[2.2]{eV/\hbar}$ \cite{kreibig}. 
Note that the data shown in Fig.\ \ref{fig:gamma_dimer}
imperceptibly differ from those in Fig.\ 3 of Ref.\ \cite{my}.

In the homogeneous dimer case [Fig.\ \ref{fig:gamma_dimer}(a)] and for the 
bright mode, the Landau damping dominates over radiation damping for 
nanoparticle sizes smaller than ca.\ $\unit[15]{nm}$ 
(compare the solid and dashed-dotted red lines in the figure). 
For the dark mode, the radiation damping strictly vanishes 
[dashed black line in Fig.\ \ref{fig:gamma_dimer}(a)], 
so that Landau damping (solid black line) is the main decay mechanism 
of the coupled plasmonic modes, until it becomes irrelevant for increasing 
nanoparticle size as compared to Ohmic losses, which are, to a first 
approximation, size-independent. 
For the Ag-Au dimer [Fig.\ \ref{fig:gamma_dimer}(b)], the dark mode acquires 
some finite dipole moment due to the difference in densities of the 
two metals, and the radiation damping of the dark mode (which is not strictly 
dark in that case) is finite and increases as a function of the nanoparticle 
size (see the dashed line in the figure). In such a case, the nonradiative 
damping is the main decay mechanism of the dark mode for nanoparticle sizes up 
to ca.\ $\unit[25]{nm}$.

\section{Nonradiative suppression of the plasmon profile along the chain for continuous and pulsed excitations}
\label{sec:details}
In this appendix, we provide details of the analytical calculations of the dipole moment on 
nanoparticle $n$ presented in Sec.\ \ref{sec:propagation}. 

The two integrals appearing in Eq.\ \eqref{eq:coefficients_calculated} are defined by
\label{sec:integrals}
\begin{equation}
\label{eq:I_n}
\mathcal{I}_n(\alpha)=\int_0^\pi\mathrm{d}x\,\frac{\sin{x}\sin{(nx)}}{\cos^2{x}+\alpha^2}
\end{equation}
and 
\begin{equation}
\label{eq:J_n}
\mathcal{J}_n(\alpha)=\int_0^\pi\mathrm{d}x\,\frac{\sin{x}\cos{x}\sin{(nx)}}{\cos^2{x}+\alpha^2},
\end{equation}
where $\alpha$ is real and positive, and $n$ is an integer strictly larger than $0$.
It is easy to show that 
\begin{equation} 
\label{eq:I_n_inter}
\mathcal{I}_n(\alpha)=\frac 12 \mathrm{Im}\int_0^{2\pi}\mathrm{d}x\,\frac{\sin{x}\,\mathrm{e}^{\mathrm{i}nx}}
{\cos^2{x}+\alpha^2}
\end{equation}
for $n$ odd and $\mathcal{I}_n(\alpha)=0$ for $n$ even, 
while
\begin{equation}
\label{eq:J_n_inter}
\mathcal{J}_n(\alpha)=\frac 12 \mathrm{Im} \int_0^{2\pi}\mathrm{d}x\,\frac{\sin{x}\cos{x}\,\mathrm{e}^{\mathrm{i}nx}}
{\cos^2{x}+\alpha^2}
\end{equation}
for $n$ even and $\mathcal{J}_n(\alpha)=0$ for $n$ odd.
Changing variables to $z=\mathrm{e}^{\mathrm{i}x}$ in Eqs.\ \eqref{eq:I_n_inter} and \eqref{eq:J_n_inter}, we arrive 
at 
\begin{equation} 
\label{eq:I_n_inter2}
\mathcal{I}_n(\alpha)=- \mathrm{Im}\oint \mathrm{d}z\,\frac{(z^2-1)z^n}
{z^4+2(1+2\alpha^2)z^2+1}
\end{equation} 
and 
\begin{equation}
\label{eq:J_n_inter2}
\mathcal{J}_n(\alpha)=- \frac 12\mathrm{Im}\oint \mathrm{d}z\,\frac{(z^4-1)z^{n-1}}
{z^4+2(1+2\alpha^2)z^2+1},
\end{equation}
where the two above integrals are taken over the unit circle in the complex plane.
The denominator of the integrands appearing in Eqs.\ \eqref{eq:I_n_inter2} and \eqref{eq:J_n_inter2}
has two simple poles lying outside of the unit circle, $z^\mathrm{out}_\pm=\pm\mathrm{i}[(1+\alpha^2)^{1/2}+\alpha]$, 
and two simple poles lying inside of the unit circle, $z^\mathrm{in}_\pm=\pm\mathrm{i}[(1+\alpha^2)^{1/2}-\alpha]$. 
By the residue theorem, we thus arrive to the final results
\begin{equation}
\label{eq:I_n_final}
\mathcal{I}_n(\alpha)=\left[1-(-1)^n\right]\mathrm{Im}\{\mathrm{i}^n\}\frac{\pi}{2}\frac{\left(\sqrt{1+\alpha^2}-\alpha\right)^n}{\alpha}
\end{equation}
and
\begin{equation}
\label{eq:J_n_final}
\mathcal{J}_n(\alpha)=-\left[1+(-1)^n\right]\mathrm{Im}\{\mathrm{i}^{n+1}\}\frac{\pi}{2}\left(\sqrt{1+\alpha^2}-\alpha\right)^n
\end{equation}
for all integers $n\geqslant1$. These simple expressions allow us to obtain the form \eqref{eq:sigma_n_analytical} of the 
dipole moment and the subsequent propagation length \eqref{eq:xi_analytical} resulting from a continuous excitation of the first nanoparticle in the chain.

The two integrals involved in the expression \eqref{eq:sigma_n_pulse_2} 
describing the dipole moment resulting from the pulsed excitation of the first nanoparticle in the chain
are defined by
\begin{equation}
\label{eq:K_def}
\mathcal{K}_n(\alpha, \beta)=\int_0^\pi\mathrm{d}x\sin{(nx)}\sin{x}\cos{(\alpha\cos{x})}\
\mathrm{e}^{-\beta\cos{x}}
\end{equation}
and 
\begin{equation}
\label{eq:L_def}
\mathcal{L}_n(\alpha, \beta)=\int_0^\pi\mathrm{d}x\sin{(nx)}\sin{x}\sin{(\alpha\cos{x})}\
\mathrm{e}^{-\beta\cos{x}}, 
\end{equation}
where $\alpha$ and $\beta$ are both real and where $n$ is an integer strictly larger than zero. 
Using that 
\begin{equation}
\int_0^\pi\mathrm{d}x\cos{(nx)}\cos{(z\cos{x})}=\pi\cos{\left(\frac{n\pi}{2}\right)}J_n(z)
\end{equation}
and 
\begin{equation}
\int_0^\pi\mathrm{d}x\cos{(nx)}\sin{(z\cos{x})}=\pi\sin{\left(\frac{n\pi}{2}\right)}J_n(z), 
\end{equation}
where $J_n(z)$ denotes the Bessel function of the first kind with $z$ a complex variable, 
we obtain 
\begin{equation}
\label{eq:K_result}
\mathcal{K}_n(\alpha, \beta)=\pi n\ \mathrm{Im}\left\{\mathrm{e}^{\mathrm{i}n\pi/2}\frac{J_n(\alpha+\mathrm{i}\beta)}{\alpha+\mathrm{i}\beta}\right\}
\end{equation}
and 
\begin{equation}
\label{eq:L_result}
\mathcal{L}_n(\alpha, \beta)=-\pi n\ \mathrm{Re}\left\{\mathrm{e}^{\mathrm{i}n\pi/2}\frac{J_n(\alpha+\mathrm{i}\beta)}{\alpha+\mathrm{i}\beta}\right\}.
\end{equation}
These closed expressions, when inserted into Eq.\ \eqref{eq:sigma_n_pulse_2} allow one
to obtain the time-dependent dipole moment $\sigma_n(t)$ and the weak-coupling 
limit \eqref{eq:sigma_finalresult} in the case of a pulsed excitation.


\end{document}